\documentclass[oupdraft]{bio}
\usepackage{url}
\usepackage{bbm}
\usepackage{pdfpages}
\usepackage{tabularx}
\usepackage{array}
\usepackage{longtable}
\usepackage{texshade}
\usepackage{graphicx,amscd,amsmath,amssymb,amsfonts,verbatim}
\usepackage{mwe}
\usepackage{mathrsfs}
\usepackage[normalem]{ulem}
\usepackage{adjustbox}
\usepackage{multirow}
\usepackage{placeins}
\usepackage{afterpage}
\usepackage{booktabs,caption}
\usepackage[flushleft]{threeparttable}
\usepackage{adjustbox}
\usepackage{rotating}
\usepackage{listings}
\usepackage{multirow}
\usepackage[margin=1in]{geometry}

\newtheorem{assumption}{Assumption}
\newtheorem{proposition}{Proposition}

\newcommand\redout{\bgroup\markoverwith
{\textcolor{red}{\rule[0.5ex]{2pt}{0.8pt}}}\ULon}

\begin{document}
  \title{Assessing the causal effects of a stochastic intervention in time series data: Are heat alerts effective in preventing deaths and hospitalizations?}
  \author{XIAO WU$^{1,4,\ast}$, 
KATE R. WEINBERGER$^{2}$, GREGORY A. WELLENIUS$^{3}$, FRANCESCA DOMINICI$^{4}$, DANIELLE BRAUN$^{4}$\\[4pt]
\textit{
$^{1}$Stanford Data Science and Department of Statistics, Stanford University, Stanford, CA 94305, USA
$^{2}$School of Population and Public Health, University of British Columbia, Vancouver, BC V6T 1Z3, Canada \\
$^{3}$Department of Environmental Health, Boston University School of Public Health, Boston, MA 02118, USA \\
$^{4}$Department of Biostatistics, Harvard T.H. Chan School of Public Health, Boston, MA 02115, USA \\
    }
% E-mail address for correspondence
$^{\ast}${wuxiao@stanford.edu}}
%\vfill

\markboth%
% First field is the short list of authors
{X. Wu and others}
% Second field is the short title of the paper
{Stochastic intervention in time series}
%\doi{10.1111/j.1541-0420.2005.00454.x}
\maketitle

\begin{abstract}
{
The methodological development of this paper is motivated by the need to address the following scientific question: does the issuance of heat alerts prevent adverse health effects? Our goal is to address this question within a causal inference framework in the context of time series data. A key challenge is that causal inference methods require the overlap assumption to hold: each unit (i.e., a day) must have a positive probability of receiving the treatment (i.e., issuing a heat alert on that day).  In our motivating example, the overlap assumption is often violated:  the probability of issuing a  heat alert on a cooler day is near zero. To overcome this challenge, we propose a  {\it  stochastic intervention} for time series data which is implemented via an incremental time-varying propensity score (ItvPS). The ItvPS intervention is executed by multiplying the probability of issuing a heat alert on day $t$ --  conditional on past information up to day $t$ --  by an odds ratio $\delta_t$. First, we introduce a new class of causal estimands that relies on the ItvPS intervention. We provide theoretical results to show that these causal estimands can be identified and estimated under a weaker version of the overlap assumption. Second, we propose nonparametric estimators based on the ItvPS and derive an upper bound for the variances of these estimators. Third, we extend this framework to multi-site time series using a spatial meta-analysis approach. Fourth, we show that the proposed estimators perform well in terms of bias and root mean squared error via simulations. Finally, we apply our proposed approach to estimate the causal effects of increasing the probability of issuing heat alerts on each warm-season day in reducing deaths and hospitalizations among Medicare enrollees in $2,837$ U.S. counties. 
}{Incremental propensity score; Meta-analysis; Multi-site time series; Time-varying confounding
}
\end{abstract}

\section{Introduction}
\label{sec:intro}
Extreme heat events are a significant threat to public health \citep{epa2006excessive}. In the U.S., heat waves have been associated with increased morbidity and mortality \citep{bobb2014cause,weinberger2020estimating}.
To reduce heat-related adverse health outcomes, the U.S. National Weather Service (NWS) issues excessive heat warnings (to indicate more severe heat events) and heat advisories (to indicate less severe heat events) in advance of forecasted heat events to communicate these risks to the public and local government officials \citep{hawkins2017assessment}. However, how effective these excessive heat warnings and advisories (collectively, ``heat alerts") are in reducing adverse health outcomes such as deaths and hospitalizations is largely unknown \citep{weinberger2021heat}.  To fill these knowledge gaps, we acquired daily time series data  during the warm months (April-October) of 2006-2016 for $N = 2,837$ U.S. counties. For each county, we obtained 1)  daily maximum heat index (an index that combines air temperature and relative humidity to posit a human-perceived equivalent temperature); 2) daily issuance of heat alerts (binary); and 3) daily number of deaths and hospitalizations among Medicare enrollees. We define the unit of analysis as {\it the day} to align with the time series literature in environmental epidemiology \citep{bell2004time}.

To analyze this data set within a causal inference framework, we need to overcome the following methodological challenges.
First, the overlap assumption \citep{rosenbaum1983central}, i.e., any day must have a positive probability of being treated (e.g., receiving a heat alert) or untreated (e.g., not receiving a heat alert), is often violated in time series data. This is because during the study period heat alerts were only issued on $2.52 \%$ of the warm-season days across U.S. counties. Moreover, a heat alert is highly unlikely on a cool day, and more likely on a very hot day. Even if the overlap assumption did hold, unrealistically large sample sizes would be needed to avoid point estimates of the causal effects with very large estimated variances \citep{kang2007demystifying}. This is because the limited overlap in the covariate distribution between treated and untreated units leads to the existence of extreme propensity scores (closed to $0$ or $1$), translating into large errors in the weights by inverse propensity weighting (IPW) methods and very low effective sample sizes for the resulting weighted populations. Second, time series observational studies are prone to time-varying confounding \citep{robins1986new}. To adjust for confounding bias, we must control for time-varying covariates (e.g., daily maximum heat index) associated with both the treatment (e.g.,  daily issuance of heat alerts) and outcomes of interest (e.g., daily deaths or hospitalizations). Third, we are dealing with multi-site time series data where it is plausible that the true causal effects of interest (e.g., whether heat alerts prevent deaths and hospitalizations in a given county) might be highly heterogeneous across counties.

Literature has focused on some aspects of the methodology gaps identified above. First, while the causal inference literature on time series studies is sparse, there are a few exceptions \citep{bojinov2019time,sobel2014causal,ning2019bayesian,papadogeorgou2020causal,shi2022dynamic}.  \cite{bojinov2019time} introduced an extended potential outcome framework for randomized experiments on time series data assuming the overlap assumption always holds. They focused on the setting where the treatment at each time can be randomly assigned based on positive probabilities. \cite{shi2022dynamic} extended this framework to online randomized experiments with reinforcement learning (see \cite{sutton2018reinforcement} for an overview). \cite{sobel2014causal,ning2019bayesian} proposed a relevant causal inference framework for time series observational studies with the focus on modeling the correlation structures of multiple time series, whereas both require the overlap assumption to be held in observational data. Therefore, their methods do not directly apply to time series observational studies with overlap violations. Second, stochastic interventions (e.g., increasing or decreasing the probability of issuing heat alerts on each day), have been proposed to overcome violations of the overlap assumption in observational studies \citep{stock1989nonparametric, robins2004effects, munoz2012population,haneuse2013estimation,kennedy2019nonparametric,imai2019comment,naimi2021incremental}. Most recently, \cite{kennedy2019nonparametric,kim2021incremental,diaz2021nonparametric} proposed causal inference framework for incremental propensity scores and modified treatment policies based on stochastic interventions, accounting for time-varying confounding in longitudinal studies. However, there are important distinctions between longitudinal studies and multi-site time series studies. More specifically, in longitudinal studies, the number of study units, which are often defined by individual patients or sites ($N$), is much larger than the number of repeated observations ($T$) for each study unit ($T << N$).  Whereas, in multi-site time series studies, the number of study units, which are instead defined by time points (e.g., days $T$), 
can be larger than the number of sites (e.g., counties $N$). Such distinctions in data structure lead to different considerations in defining causal estimands and developing statistical methods.
\cite{papadogeorgou2020causal} are the first to bridge a stochastic intervention with time series observational data, but they focused on treatments that can be modeled  by a spatio-temporal point process. In this context, they analyzed a single time series that contains information on the geographic coordinates of treatments (e.g., airstrikes).  In the presence of heterogeneity of the true causal effects across counties, as is the case in our motivating example, the causal estimands defined on a homogeneous super-population -- as is done in the context of longitudinal studies or for a spatio-temporal point process --  might not capture the heterogeneous nature of the study population. 

To the best of our knowledge, a causal inference approach in the context of a stochastic intervention for multi-site time series is lacking. Accordingly, in this paper, we introduce a stochastic intervention defined by the incremental time-varying propensity score (ItvPS) for time series data (i.e., the ItvPS intervention). The ItvPS intervention is executed by multiplying the probability of receiving the treatment at time $t$ conditional on past information up to time $t$ by an odds ratio $\delta_t$ (e.g., increase the probability of issuing heat alerts on day $t$). This novel framework allows us to define a broad class of stochastic causal estimands on time series data and ease the identification and estimation due to violation of the overlap assumption. 

In Section~\ref{sec:setup}, we set up the notations in the context of our motivating example about national data on heat alerts. In Section~\ref{sec:meth}, starting with a single time series, we define the stochastic causal estimands based on ItvPS intervention and their corresponding assumptions of identifiability. Next, in Section~\ref{estimation}, we propose nonparametric estimators based on ItvPS, derive an upper bound for the variances of these estimators, and construct the corresponding asymptotic confidence intervals (CIs) and time-uniform confidence sequences (CSs). In Section~\ref{sec:meta}, we extend our approach into the context of multi-site time series, introduce additional assumptions of identifiability for multi-site time series, and propose a random-effects meta-analysis to pool causal estimands across time series from multiple sites. In Section~\ref{sec:sim}, we illustrate the finite-sample performance of the proposed estimators via simulation studies. In Section~\ref{sec:app}, we apply our method to the national heat alert data set to estimate the effectiveness of heat alerts in reducing morbidity and mortality among Medicare beneficiaries. In Section~\ref{sec:conc}, we conclude with a summary and discussion.

\section{Set up}
\label{sec:setup}
\subsection{Mathematical Notations}
\label{notation}
We introduce the following notation in the context of time series data. Let $Y_{i,t}, W_{i,t}, \mathbf{C}_{i,t}$ be the outcome, treatment and pre-treatment covariates at time $t$, respectively, for $t \in \{1,2,\ldots,T\}$ in site $i$, for $i \in \{1,2,\ldots,N\}$.
First, we introduce the causal estimands in the context of a single time series and omit the index $i$ for conciseness. Later, in Section~\ref{sec:meta}, we reintroduce the index $i$ to indicate multi-site time series.

For a single time series, each time $t$ can be assigned to the treatment $W_{t} = 1$ or $W_{t} = 0$, and then the outcome $Y_{t}$ is observed. We assume that $W_t$ is  binary; $Y_t$ and $\mathbf{C}_t$ can be binary, categorical, or continuous. We define  $\{\mathbf{C}_{1:T} = (\mathbf{C}_{1},...,\mathbf{C}_{T}), {W}_{1:T} = (W_{1},...,W_{T}), {Y}_{1:T} = (Y_{1},...,Y_{T})\}$. We denote by  $\mathcal{F}_t$  the filtration which captures past information  prior to the treatment assignment at time $t$, that is $\mathcal{F}_t = \{ \mathbf{C}_{1:t}, {W}_{1:(t-1)}, {Y}_{1:(t-1)}\}$ (note, $\mathbf{C}_t$ included in the filtration $\mathcal{F}_t$ because it precedes the treatment that takes place at time $t$). Following the potential outcome framework for time series data  \citep{bojinov2019time}, we denote $Y_t({w}_{1:t})$ the potential outcome at time $t$ that would have been observed under the treatment path ${w}_{1:t} =(w_1,...,w_t)$, and we introduce the potential outcome path $Y_{1:t}({w}_{1:t}) = \{Y_1(w_{1:1}),Y_2(w_{1:2}),...,,Y_t(w_{1:t})\}$. 

We define the time-varying propensity score at time $t$ as
$p_t(w_t,\mathcal{F}_t) = Pr(W_t = w_t | \mathcal{F}_t), \ \text{for}\ w_{t} = \{0,1\}.$
In our motivating example, the time-varying propensity score for treatment $W_t = 1$ denotes the probability of issuing a heat alert on day $t$ conditional on past information up to day $t$ prior to the heat alert issuance. We extend the incremental propensity score interventions proposed by \cite{kennedy2019nonparametric} to the time series studies. The incremental propensity score intervention is an intervention that shifts the value of the propensity score. This is in contrast with the traditional deterministic intervention which instead is implemented by shifting the value of treatment (See Section~\ref{motivate_data} for an example). First, we introduce the notation of the ItvPS as:
 \begin{align}
p^{\text{ItvPS}}_t(w_t,\mathcal{F}_t) := \frac{w_t \delta_t p_t(1,\mathcal{F}_t) + (1 - w_t) p_t(0,\mathcal{F}_t) }{\delta_t p_t(1,\mathcal{F}_t) + p_t(0,\mathcal{F}_t)}, \ \text{for}\ w_{t} = \{0,1\}.
\label{ItvPS}
    \end{align}
Note that $p^{ItvPS}_t(w_t, \mathcal{F}_t)$ is obtained by solving the following Equation
\begin{align}
    \delta_t = \frac{p^{\text{ItvPS}}_t(1,\mathcal{F}_t)/p^{\text{ItvPS}}_t(0,\mathcal{F}_t)}{p_t(1,\mathcal{F}_t)/p_t(0,\mathcal{F}_t)},
\label{delta_n}
\end{align}
where the odds ratio $\delta_t$ is set \textit{a priori} by the analyst.
Next, we define a stochastic intervention by the notation of ItvPS, and call it an ItvPS intervention. By executing an ItvPS intervention, we replace the probability of receiving the treatment $p_t(1,\mathcal{F}_t)$ at time $t$ by $p^{\text{ItvPS}}_t(1,\mathcal{F}_t)$. Such a stochastic intervention reflects an odds ratio change of $\delta_t$ in the time-varying propensity scores.
In the policy-relevant question, this can be explained as multiplying the probability of receiving the treatment on day $t$ -- conditional on past information up to day $t$, $\mathcal{F}_t$ -- by an odds ratio $\delta_t$ (see Equation~\ref{delta_n}).

In the context of a time series with length $T$, we denote $\delta_{1:T} = \{\delta_1,\delta_2,...,\delta_T\}$ as the whole intervention path and $w_{1:T}^{\text{ItvPS}}(\delta_{1:T})$ as the post-intervened treatment path under the intervention path $\delta_{1:T}$.

\vspace{-0.5cm}
\subsection{Motivating Example}
\label{motivate_data}
To illustrate these mathematical notations in our motivating example, the binary treatment $W_t$ indicates the issuance of a heat alert on day $t$ in a given county. The covariates $\mathbf{C}_t$ include (forecasted) population-weighted daily maximum heat index that occurred just prior to treatment on day $t$, day of the week, and federal holidays which may be related to both heat alert issuance and adverse health outcomes \citep{weinberger2021heat}. The outcome $Y_t$ denotes the daily number of all-cause deaths or cause-specific hospitalizations for disease causes that were found to be associated with extreme heat in the Medicare population \citep{bobb2014cause}. Note, we do not have information on cause-specific deaths, therefore we focus on all-cause deaths. The filtration $\mathcal{F}_t$ represents the past information prior to the treatment assignment on day $t$. The inclusion of past treatments $W_{1:t-1}$ and/or outcomes $Y_{1:t-1}$ is also allowed through the filtration, aligning with other literature \citep{bojinov2019time,kennedy2019nonparametric}.

While the exact criteria used to issue heat alerts varies across the jurisdictions of local NWS offices, a key commonality across jurisdictions is that the issuance of heat alerts is based on forecasts of future weather conditions based on  past information. Therefore, the probability of issuing a heat alert on day $t$, $p_t(w_t,\mathcal{F}_t)$ is modeled based on the past information $\mathcal{F}_t$ as discussed in the previous paragraph. 

\begin{figure}[ht]
\centering
\includegraphics[width=1\linewidth]{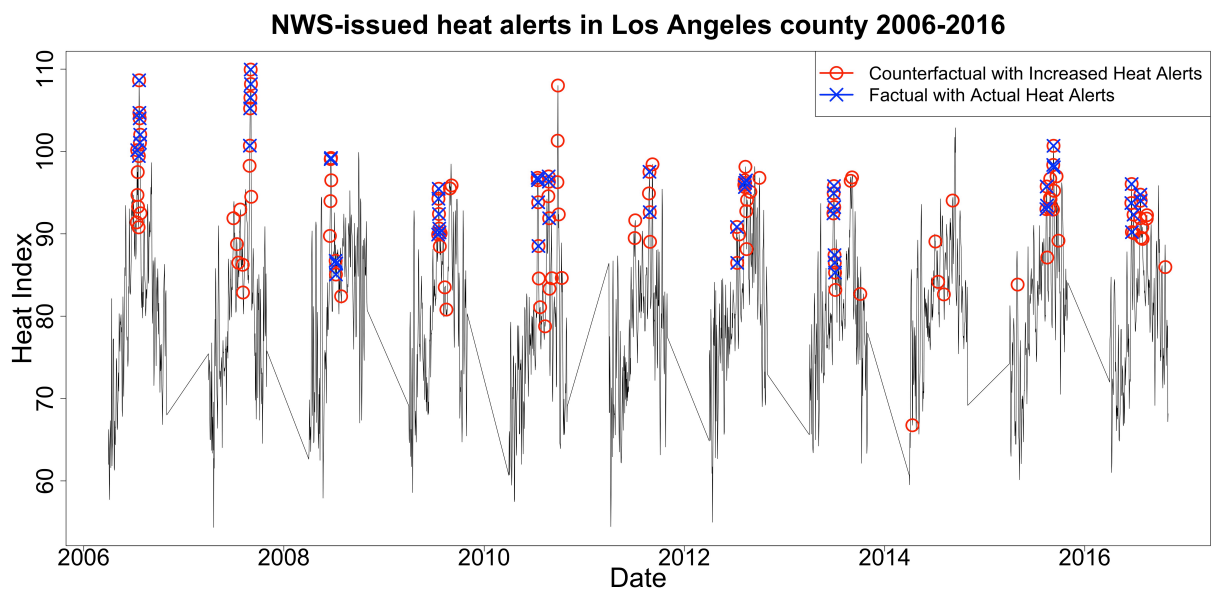}
\caption{NWS-issued heat alerts for Los Angeles, CA during the warm months (April-October) of 2006-2016. The blue x's represent the heat alerts that were issued \textit{factually}, and the red circles represent the anticipated heat alerts under a counterfactual scenario where the probability of issuing heat alerts was increased for every warm-season day by an odds ratio of $10$. Therefore, we find under this counterfactual scenario, Los Angeles, CA, would have issued $128$ heat alerts from 2006-2016, compared to $56$ heat alerts that were actually issued.
}
\label{la_heat}
\end{figure}

We consider a hypothetical ItvPS intervention in Los Angeles, CA, where one could increase the probability of issuing heat alerts by an odds ratio $\delta_t = 10$ for every day $t$ during the warm months (April-October) of 2006-2016. Figure~\ref{la_heat} illustrates the comparison between the heat alerts that were issued and the heat alerts that could have been issued under this ItvPS intervention. For example, if the probability of issuing a heat alert on day $t$ is 90\%, under the incremental propensity score intervention with an odds ratio $\delta_t=10$, the probability of issuing a heat alert would increase from 90\% to $\frac{10\times 90\%}{10 \times 90\% +1- 90\%} \approx 98.9\%$. Note that increasing the probability of issuing a heat alert by an odds ratio of $10$ for each day is not equivalent to increasing the total number of heat alerts by $10$ times. We find that if we hypothetically increase the probability of issuing a heat alert by an odds ratio of $10$ on each warm-season days, the NWS office would have issued $128$ heat alerts (an average of approximately $12$ alerts per year) in the warm months of 2006-2016, compared to the total of $56$ heat alerts (an average of approximately $5$ alerts per year) actually issued.

\section{Causal Estimand and Identification}
\label{sec:meth}
\subsection{General Estimand}
\label{general}
We focus on the following scientific question ``If we had changed the probability of issuing heat alerts by some pre-specified amount on each day in the study period, how many total deaths would have been averted and how many cause-specific hospitalizations for heat-related diseases could have been avoided?" This question can be answered by quantifying the causal effect of an ItvPS intervention. We first define the general causal estimand of an ItvPS intervention in the context of time series data. The parameter $\tau_t(\delta_{1:t}):= E[Y_t\{w_{1:t}^{\text{ItvPS}}(\delta_{1:t})\}]$ denotes the mean potential outcome at time $t$ with respect to the post-intervened treatment path $w_{1:t}^{\text{ItvPS}}(\delta_{1:t})$. In econometrics, this parameter belongs to the class of dynamic causal effects since the parameter changes over time \citep{rambachan2021common}. In our motivating example, this would be the counterfactual daily deaths or hospitalizations on day $t$ given the ItvPS intervention. 

In cross-sectional or longitudinal studies, causal estimands are generally defined as the average of potential outcomes across $N$ patients or sites. In the context of a single time series, we define the  causal estimand as the temporal average of the potential outcomes up to time $T$.  Under the post-intervened treatment path $w_{1:t}^{\text{ItvPS}}(\delta_{1:t})$ where the probability of receiving the treatment on day $t$ has been multiplied by an odds ratio $\delta_t$ for $t = 1,\ldots, T$, we define following temporal-average causal estimand
$
     \bar{{\tau}} (\delta_{1:t,T}) = \frac{1}{T}\sum^T_{t=1} {\tau}_{t} (\delta_{1:t}).
$
In our motivating example, for Los Angeles, CA (see Figure~\ref{la_heat}), the parameter $\bar{{\tau}} (\delta_{1:t,T})$  denotes  the average number of deaths or cause-specific hospitalizations per day  under the hypothetical scenario where the  probability of issuing a heat alert is multiplied by an odds ratio $\delta_t = 10$ for each day $t = 1,\ldots,T$ during the warm season.

\subsection{Causal Estimand on Observed Treatment Path}
\label{t0_estimand}
Without further modeling assumptions, estimating the causal estimand $\bar{{\tau}} (\delta_{1:t,T})$ defined in Section~\ref{general} might be challenging for large $T$ because this parameter depends  on a treatment path with length $T$. Building upon \cite{bojinov2019time}, we define a duration-$t_0$ causal estimand conditional on the observed treatment path $w^{obs}_{1:(t-t_0-1)}$ denoted as $\tau_t(\delta_{(t-t_0):t})$ (i.e., the duration of interventions is up to $t_0$ days), and its temporal average $ \bar{\tau}(\delta_{(t-t_0):t,T})$, as follows
\begin{align*}
   \tau_t(\delta_{(t-t_0):t}) = E[Y_t\{w^{obs}_{1:(t-t_0-1)}, w_{(t-t_0):t}^{\text{ItvPS}}(\delta_{(t-t_0):t})\}], \ t_0=0,\ldots,t-1; \
   \bar{\tau}(\delta_{(t-t_0):t,T}) = \frac{1}{T-t_0} \sum^T_{t=t_0+1} \{\tau_t(\delta_{(t-t_0):t})\}.
\end{align*}
Defining estimands dependent on pre-specified fixed memory is common in time series modeling, e.g., in an autoregressive integrated moving average (ARIMA) model, the fixed memory up to lagged $t_0$ data points was pre-specified. They are also compatible with the causal inference literature since they can be treated as conditional causal estimands conditioning on partial historical information $\{W_{1:(t-t_0-1)} = w^{obs}_{1:(t-t_0-1)}\}$ \citep{imbens2015causal}. Related theoretical developments were studied by \cite{van2018robust}, studying a class of conditional (or named ``context-specific") causal parameters for a single time series, conditioning on a fixed dimensional summary measure of the historical information. Furthermore, they established the statistical properties of their corresponding estimators under a targeted maximum likelihood estimation (TMLE) framework. Along this line, we define the causal estimands dependent on the observed treatment path $w^{obs}_{1:(t-t_0-1)}$ as a class of conditional causal estimands conditioning on historical information. We later show the estimands can be estimated using a weighted estimation strategy (see Section~\ref{estimation}). 

\subsection{Assumptions and Identification}
Following the  potential outcomes framework for time series data \citep{bojinov2019time}, we introduce the following assumptions of identifiability:
\begin{assumption}[SUTVA]
The outcome path satisfies non-anticipating, consistency, and non-interference assumptions, that is, 
\label{consistency}
$
     Y^{\text{obs}}_{t} = Y_{t}(w^{\text{obs}}_{1:T}) = Y_{t}(w^{\text{obs}}_{1:t}) \ \forall \ t = 1,...,T.
$
 \end{assumption}
We use the stable unit treatment value assumption (SUTVA) to articulate the three conditions that are required for the potential outcomes framework for time series data \citep{vanderweele2013causal,bojinov2019time}.
1) The potential outcomes at time $t$ can depend on the treatment path  up to time $t$, but are not allowed to depend on future treatments (non-anticipating). 2) There is only one version of the treatment for each time $t$, and each treatment path up to time $t$ realizes a unique observed outcome at time $t$ (consistency). 3) Since we only have one site, there is no spillover effect from other sites (non-interference).

\begin{assumption}[Unconfoundedness]
\label{unconfounded}
The assignment mechanism is unconfounded if for all $W_{1:T} \in \mathcal{W} = \{0,1\}^T$, 
$
W_t \ \perp \ Y_s(w_{1:s}) \ \mid \ \mathcal{F}_t \ \forall \ t = 1,...,T \ \text{and} \ t \leq s \leq T.
$
\end{assumption}
Assumption~\ref{unconfounded} aligns with the ``sequential randomization" assumption  introduced in longitudinal studies by \cite{robins1994estimation}, which states that the treatment assignment only depends on past information and thus is conditionally independent of future potential outcomes. This assumption also excludes the possibility that  future potential outcomes could impact the current treatment assignment retrospectively (a phenomenon that has been discussed in \cite{granger1980testing}).

\begin{assumption}[Weak overlap] 
The assignment mechanism  weakly overlaps if, for all $t \in \{1,2,...,T\}$, there exists a constant $\gamma >0$, such that
$
p^{\text{ItvPS}}_t(w_t, \mathcal{F}_t) \leq \gamma p_t(w_t, \mathcal{F}_t),
$ $\forall w_t \in \{0,1\}$.
\end{assumption}
The overlap assumption in traditional causal inference literature requires that each time $t$ has a positive probability of being treated or untreated. In contrast, a stochastic intervention framework may avoid the overlap assumption by not intervening at time points with zero probability of being either treated or untreated \citep{kennedy2019nonparametric}. Instead, the stochastic intervention framework only requires a weak overlap assumption to hold \citep{papadogeorgou2020causal}. Specifically, the weak overlap assumption always holds under the proposed ItvPS intervention (a special type of stochastic intervention), since, according to Equation~\ref{ItvPS}, $p^{\text{ItvPS}}_t(w_t, \mathcal{F}_t) \leq \max\{1, \delta_t\} p_t(w_t , \mathcal{F}_t)$, and $p^{\text{ItvPS}}_t(1, \mathcal{F}_t) \equiv 0$ when $p_t(1, \mathcal{F}_t) = 0$ and $p^{\text{ItvPS}}_t(0, \mathcal{F}_t) \equiv 0$ when $p_t(0, \mathcal{F}_t) = 0$ regardless of the value of $\delta_t$ \citep{naimi2021incremental}.

Under the assumptions defined above, our proposed causal estimand is identified: at any time point $t$, 
    \begin{align*}
         {\tau}_t(\delta_{1:t}) &= \sum_{w_{1:t} \in W_{1:t}}  \int_{\partial \mathcal{R}_{1:t}} \mu(w_t, \mathcal{F}_{t}) \times \prod^t_{s=1} \underbrace{ \big[ \frac{w_{s} \delta_s p_s(1, \mathcal{F}_{s})  + (1-w_{s})   p_s(0,\mathcal{F}_{s}) }{ 
     \delta_s {p_s(1,\mathcal{F}_{s})} +  p_s(0,\mathcal{F}_{s})} \big]}_{:=p^{\text{ItvPS}}_s(w_s,\mathcal{F}_{s})} \times d Pr(\partial \textit{r}_{s}  | W_{s-1} = w_{s-1}, \mathcal{F}_{s-1}),
    \end{align*}
where $\partial \mathcal{R}_{1:t} = \partial \mathcal{R}_{1} \times ... \times \partial \mathcal{R}_{t}$, $\partial \mathcal{R}_{s} = \mathcal{F}_{s}/ \{\mathcal{F}_{s-1},W_{s-1} \},s=1,...,t$, and $\mu(w_t, \mathcal{F}_{t}) = E(Y_t \mid W_t = w_t, \mathcal{F}_{t}), t = 1,...,T$. The proof is provided in the Supplementary Materials Section S.1. However, although $\tau_t(\delta_{1:t})$ is causally identifiable, the corresponding quantity may not be stably estimated using a single time series. We instead focus on the temporal average causal estimand
$
     \bar{{\tau}} (\delta_{1:t,T}) = \frac{1}{T}\sum^T_{t=1} {\tau}_{t} (\delta_{1:t}),
$
which can also be causally identified given  ${\tau}_{t} (\delta_{1:t})$ is identifiable $\forall t = 1,2,...,T$. Likewise, the duration-$t_0$ causal estimand $\tau_t(\delta_{(t-t_0):t})$ and its temporal average $\bar{\tau}(\delta_{(t-t_0):t,T})$ can be identified under the same assumptions.

\section{Estimation and Inference}
\label{estimation}
\subsection{Weighted Estimation}
We focus on the estimation and inference of duration-$t_0$ causal estimands $\tau_t(\delta_{(t-t_0):t})$ given the theoretical and practical conveniences discussed in Section~\ref{t0_estimand}.
 If the value of time-varying propensity score $p_s(w_s,\mathcal{F}_{s})$ is known, as in Section 6 of \cite{kim2021incremental}, the  unbiased estimator of $\tau_t(\delta_{(t-t_0):t})$ is defined as,
\begin{align*}
        \hat{{\tau}}_t(\delta_{(t-t_0):t}) &= \prod^t_{s=t-t_0}  \big[\underbrace{\frac{\{W_{s} \delta_s   + (1-W_{s}) \} }{ 
     \delta_s {p}_s(1,\mathcal{F}_{s}) +  {p}_s(0,\mathcal{F}_{s})}}_{:={p_s^{\text{ItvPS}}}(W_s,\mathcal{F}_{s})/p_s(W_s,\mathcal{F}_{s})} \big] Y_t.
\end{align*}
When the time-varying propensity score  $p_s(w_s,\mathcal{F}_{s})$ is unknown, the estimation strategy requires two steps: 1) at each time $s$, we estimate the time-varying propensity score  $\hat{p}_s(w_s,\mathcal{F}_{s})$; 2) if the time-varying propensity scores can be modeled using correctly specified parametric models, we construct the following unbiased estimator for a duration-$t_0$ estimand on the observed treatment path,
    \begin{align*}
        \hat{{\tau}}_t(\delta_{(t-t_0):t}) &= \prod^t_{s=t-t_0}  \big[\underbrace{\frac{\{W_{s} \delta_s   + (1-W_{s}) \} }{ 
     \delta_s \hat{p}_s(1, \mathcal{F}_{s}) + \hat{p}_s(0, \mathcal{F}_{s})}}_{:=\widehat{p}^{\text{ItvPS}}_s(w_s,\mathcal{F}_{s})/\widehat{p}_s(w_s,\mathcal{F}_{s})} \big] Y_t ,\\
         \hat{\bar{{\tau}}} (\delta_{(t-t_0):t,T}) &= \frac{1}{T-t_0}\sum^T_{t=t_0+1} \hat{\tau}_{t} (\delta_{(t-t_0):t})
         = \frac{1}{T-t_0} \sum^T_{t=t_0+1} \Big( \prod^t_{s=t-t_0} \big[ \frac{\{W_{s} \delta_s   + (1-W_{s}) \} }{ 
     \delta_s \hat{p}_s(1, \mathcal{F}_{s}) + \hat{p}_s(0, \mathcal{F}_{s})} \big]Y_t \Big).
    \end{align*}
The proposed estimator is defined as a weighted average of the observed outcomes $Y_t$ weighted by time-varying weights, in which  the weights correspond to the product of fractions where the numerators are $\widehat{p}^{\text{ItvPS}}_s(w_s,\mathcal{F}_{s})$ and denominators are $\widehat{p}_s(w_s,\mathcal{F}_{s})$. All estimators described above are generally referred to as IPW estimators, since they take estimates for the time-varying propensity score  $\hat{p}_s(w_s,\mathcal{F}_{s})$, and plug it into the denominator of the weights.

\subsection{Statistical Inference}
We investigate the variance of the proposed IPW estimators.
For a fixed $t$, conditioning on $\mathcal{F}_{t-t_0}$, the variance of the estimator $\hat{{\tau}}_t(\delta_{(t-t_0):t})$ can be defined as
    \begin{align*}
    \small
       Var\{\hat{{\tau}}_t(\delta_{(t-t_0):t})| \mathcal{F}_{t-t_0}\} = \underbrace{E\Big(  \prod^t_{s=t-t_0} \big[\frac{ \{ W_{s} \delta_s   + (1-W_{s}) \} }{ 
     \delta_s \hat{p}_s(1, \mathcal{F}_{s}) + \hat{p}_s(0, \mathcal{F}_{s})} \big]^2 Y_t^2 \Big| \mathcal{F}_{t-t_0} \Big)}_{\mathcal{V}_t} - \Big\{ E \Big( \prod^t_{s=t-t_0}  \big[\frac{\{W_{s} \delta_s   + (1-W_{s}) \} }{ 
     \delta_s \hat{p}_s(1, \mathcal{F}_{s}) + \hat{p}_s(0, \mathcal{F}_{s})} \big] Y_t \Big| \mathcal{F}_{t-t_0} \Big)\Big\}^2,
\end{align*}
where  $\mathcal{V}_t$ is defined as
\begin{align*}
     \mathcal{V}_t &= \sum_{w_{(t-t_0):t} \in W_{(t-t_0):t}}  \int_{\partial \mathcal{R}_{(t-t_0):t}} E(Y_t^2 \mid W_{(t-t_0):t} = w_{(t-t_0):t}, \mathcal{F}_{t}) \times  \\ & \ \ \ \  \prod^t_{s=t-t_0} \big[ \frac{\{w_{s} \delta_s p_s(1, \mathcal{F}_{s})  + (1-w_{s})  p_s(0,\mathcal{F}_{s})\} }{ 
     \delta_s p_s(1, \mathcal{F}_{s}) +  p_s(0, \mathcal{F}_{s})}\big]^2 \times d Pr(\partial {r}_{s}  | W_{s-1} = w_{s-1}, \mathcal{F}_{s-1}).
     \end{align*}
The IPW estimator of $\mathcal{V}_t$ can be written as
\begin{align}
     \hat{\mathcal{V}}_t &= \prod^t_{s=t-t_0}  \big[\frac{\{W_{s} \delta_s^2   + (1-W_{s}) \} }{ 
     [\delta_s \hat{p}_s(1, \mathcal{F}_{s}) +  \hat{p}_s(0, \mathcal{F}_{s})]^2} \big] Y_t^2.
          \label{upper_var}
    \end{align}
A straightforward IPW estimator of $Var\{\hat{{\tau}}_t(\delta_{(t-t_0):t})| \mathcal{F}_{t-t_0}\}$ can be written as
\begin{align*}
  \widehat{Var}\{\hat{{\tau}}_t(\delta_{(t-t_0):t})| \mathcal{F}_{t-t_0}\} &=  \prod^t_{s=t-t_0}  \big[\frac{\{W_{s} \delta_s^2   + (1-W_{s}) \} }{ 
     [\delta_s \hat{p}_s(1, \mathcal{F}_{s}) +  \hat{p}_s(0, \mathcal{F}_{s})]^2} \big] Y_t^2 - \big(\prod^t_{s=t-t_0}  \big[\frac{\{W_{s} \delta_s   + (1-W_{s}) \} }{ 
     \delta_s \hat{p}_s(1, \mathcal{F}_{s}) +  \hat{p}_s(0, \mathcal{F}_{s})} \big] Y_t \big)^2.
\end{align*}
Note, this variance estimator on a single time series is always equal to $0$, i.e.,
\begin{align*}
& \ \ \ \ \prod^t_{s=t-t_0}  \big[\frac{\{W_{s} \delta_s^2   + (1-W_{s}) \} }{ [\delta_s \hat{p}_s(1, \mathcal{F}_{s}) +  \hat{p}_s(0, \mathcal{F}_{s})]^2} \big] Y_t^2 - \big(\prod^t_{s=t-t_0}  \big[\frac{\{W_{s} \delta_s   + (1-W_{s}) \} }{ 
     \delta_s \hat{p}_s(1, \mathcal{F}_{s}) +  \hat{p}_s(0, \mathcal{F}_{s})} \big] Y_t \big)^2 \\
&=   \big[\frac{ \prod^t_{s =t-t_0 \& W_s = 1} \delta_s^2 }{ \prod^t_{s=t-t_0} \{\delta_s \hat{p}_s(1, \mathcal{F}_{s}) +  \hat{p}_s(0, \mathcal{F}_{s})\}^2} \big] Y_t^2 - \big(  \big[\frac{  \prod^t_{s =t-t_0 \& W_s = 1} \delta_s}{ \prod^t_{s=t-t_0}
   \{  \delta_s \hat{p}_s(1, \mathcal{F}_{s}) +  \hat{p}_s(0, \mathcal{F}_{s}) \}} \big]^2 Y_t^2 \big) = 0.
\end{align*}
Therefore, instead, following the time series literature \citep{bojinov2019time,papadogeorgou2020causal}, we use ${\mathcal{V}_t}$ as an upper bound of the conditional variance of $\hat{{\tau}}_t(\delta_{(t-t_0):t})$ conditioning on $\mathcal{F}_{t-t_0}$. This is because we have
\begin{align*}
  {Var}\{\hat{{\tau}}_t(\delta_{(t-t_0):t})| \mathcal{F}_{t-t_0}\} = {\mathcal{V}_t} - \Big\{ E \Big( \prod^t_{s=t-t_0}  \big[\frac{\{W_{s} \delta_s   + (1-W_{s}) \} }{ 
     \delta_s \hat{p}_s(1, \mathcal{F}_{s}) +  \hat{p}_s(0, \mathcal{F}_{s})} \big] Y_t  \Big| \mathcal{F}_{t-t_0} \Big)\Big\}^2  \leq   {\mathcal{V}_t}, 
\end{align*}
and the IPW estimator  $\hat{\mathcal{V}}_t$ can be directly calculated using observed data as shown in Equation~\ref{upper_var}. 

To calculate the variance of the temporal average estimator $\hat{\bar{{\tau}}} (\delta_{(t-t_0):t,T}) = \frac{1}{T-t_0}\sum^T_{t=t_0+1} \hat{\tau}_{t} (\delta_{(t-t_0):t})$, we firstly define a sequence for the estimation error $u_{t,t_0} \equiv \hat{\tau}_t (\delta_{(t-t_0):t}) - \tau_t (\delta_{(t-t_0):t}), \forall t = 1,2,\ldots,T$ ($t_0$ is fixed). And then, we have
$
   Var\{ \hat{\bar{{\tau}}} (\delta_{(t-t_0):t,T})\} = 
   Var\{ \frac{1}{T-t_0} \sum^T_{t=t_0+1} u_{t,t_0} \}.
$
Analogous to the proof of Theorem 1 in \cite{bojinov2019time} and Lemma 1 in \cite{papadogeorgou2020causal}, $u_{t,t_0}$ is a martingale difference sequence with respect to $\mathcal{F}_{t-t_0}$ given $u_{t,t_0}$ is bounded and $E(u_{t,t_0} | \mathcal{F}_{t-t_0}) = 0$. Consequently, the sequence $u_{t,t_0}$ is uncorrelated through time $t$, and thus
\begin{align*}
   Var\{ \hat{\bar{{\tau}}} (\delta_{(t-t_0):t,T})\} &= \frac{1}{(T-t_0)^2} \sum^T_{t=t_0+1} Var \{u_{t,t_0} \} = \frac{1}{(T-t_0)^2} \sum^T_{t=t_0+1} E[Var \{u_{t,t_0}  | \mathcal{F}_{t-t_0} \}] \\
   &= \frac{1}{(T-t_0)^2} \sum^T_{t=t_0+1} E[Var \{\hat{\tau}_{t} (\delta_{(t-t_0):t}) | \mathcal{F}_{t-t_0} \}]
   \leq \frac{1}{(T-t_0)^2} \sum^T_{t=t_0+1} {\mathcal{V}_t},
\end{align*}
and the IPW estimator of an upper bound of  $Var\{ \hat{\bar{{\tau}}} (\delta_{(t-t_0):t,T})\}$ can be expressed as $\frac{1}{(T-t_0)^2} \sum^T_{t=t_0+1} \hat{\mathcal{V}}_t$.

Using the martingale sequence property for the estimation error $u_{t,t_0}$, we can establish a central limit theorem (CLT) of the proposed estimators $\hat{\bar{{\tau}}} (\delta_{(t-t_0):t,T})$ that allows us to build CIs.
\begin{proposition}[Asymptotic Normality] 
   Suppose that Assumptions $1-3$ hold, if $\frac{1}{T-t_0} \sum^T_{t=t_0+1} E(u_{t,t_0}^2 | \mathcal{F}_{t-t_0}) \overset{p}{\longrightarrow} V^{*}$ for a positive constant $V^{*}$, then as $T \rightarrow \infty$, for a given $t_0$ we have,
    $
        \sqrt{T-t_0} \big\{ \hat{\bar{{\tau}}} (\delta_{(t-t_0):t,T}) - \bar{{\tau}} (\delta_{(t-t_0):t,T}) \big\} =
        \sqrt{T-t_0} \big\{ \frac{1}{T-t_0} \sum^T_{t=t_0+1} u_{t,t_0} \big\}
       \overset{d}{\longrightarrow} \mathcal{N}(0, V^{*}),
    $
    where $V^{*} = \lim_{T \rightarrow \infty}\frac{1}{(T-t_0)} \sum^{T}_{t=t_0+1} Var\{u_{t,t_0}| \mathcal{F}_{t-t_0}\}$.
\end{proposition}
The proof is provided in the Supplementary Materials Section S.2. Here, we note that an upper bound of $V^{*}$ is $\lim_{T \rightarrow \infty} \frac{1}{(T-t_0)} \sum^{T}_{t=t_0+1} {\mathcal{V}}_t$. Then, we construct the point-wise Wald $100(1-\alpha)$\% CI of $\hat{\bar{{\tau}}} (\delta_{(t-t_0):t,T})$ as $\{\hat{\bar{{\tau}}} (\delta_{(t-t_0):t,T}) \pm z_{1-\alpha/2} \times \sqrt{\frac{1}{(T-t_0)^2} \sum^T_{t=t_0+1} \hat{\mathcal{V}}_t}\}$, where $z_{1-\alpha/2}$ denotes the upper $\alpha/2$ critical value of a standard normal distribution.

We also construct a time-uniform CS for the parameter $\bar{\tau}(\delta_{(t-t_0):t,T})$. The $100(1-\alpha)$\% CS is a sequence of CIs $(L_T, U_T)$ that are constructed from the first $T$ samples, and have a uniform (simultaneous) coverage guarantee \citep{darling1967confidence}, i.e.,
$
        {P}\{\forall T \ge t_0 + 1: \bar{\tau}(\delta_{(t-t_0):t,T}) \in (L_T, U_T) \}  \ge 1- \alpha.
$
We yield the following $100(1-\alpha)$\% CS for $\bar{\tau}(\delta_{(t-t_0):t,T})$ based on Theorem 2 of \cite{waudby2021time}: Setting $\hat{\bar{V}}_T = \frac{1}{T-t_0} \sum^T_{t=t_0+1}\hat{\mathcal{V}}_t$, for any pre-specified constant $\rho > 0$, the $100(1-\alpha)$\% Lypaunov-type asymptotic CS for $\bar{\tau}(\delta_{(t-t_0):t,T})$ is 
$
        \Big\{\hat{\bar{{\tau}}} (\delta_{(t-t_0):t,T}) \pm \sqrt{\frac{2\{(T-t_0) \hat{\bar{V}}_T \rho^2(T) +1\}}{(T-t_0)^2 \rho^2(T)} \log\Big(\frac{\sqrt{(T-t_0) \hat{\bar{V}}_T \rho^2(T)+1}}{\alpha}\Big)}\Big\},
$
    with an approximate solution of time-dependent $
        \rho(T) := \sqrt{\frac{-\alpha^2-2\log\alpha+\log(-2\log\alpha+1-\alpha^2)}{T-t_0}}
    $ to optimize the boundary of CS.

\vspace{-0.5cm}
\section{Meta-analysis on Multi-site Time Series}
\label{sec:meta}
In many applications, time series data are available from multiple sites. In this section, we generalize our ItvPS intervention framework to combine information from  multi-site time series. To formalize the causal identification in the context of random-effects across multi-site time series, in Section~\ref{assum_multi} we introduce and discuss new assumptions of identification. In subsequent Section~\ref{inf_multi}, we describe the estimation and inference of the pooled estimator based on meta-analysis models.

\subsection{Assumptions and Identification}
\label{assum_multi}
There are at least two popular statistical models for meta-analysis, the fixed-effect model and the random-effects model \citep{borenstein2010basic}. The fixed-effect model assumes that the underlying true effect is homogeneous for different sites, i.e., there is a common effect across all sites. The random-effects model allows effect heterogeneity across sites, whereas assumes the effect in each site is treated as a random sample from a super-population.
In general, we found the random-effects assumption is more plausible in many applications including our motivating example, allowing us to assume that the true causal effects of heat alerts on health outcomes vary across counties. Therefore, we propose a meta-analysis method that relies on a random-effects model to obtain a pooled estimator to summarize the overall causal effect of time series data across multiple counties \citep{dersimonian2015meta}.

Before introducing the new pooled estimator, we introduce additional assumptions to allow for causal inference that synthesizes evidence from multi-site time series. We use the index $i$ to indicate the sites (i.e., counties in our motivating example), for $i \in \{1,2,\ldots,N\}$. We denote by $\boldsymbol{\mathcal{F}}_t = \{\mathcal{F}_{1,t},...,\mathcal{F}_{N,t}\}$ the filtration which captures the past information prior to time $t$ in $N$ sites.

\begin{assumption}[Multi-site Time Series SUTVA]
The multi-site outcome paths satisfy non-anticipating, consistency, and non-interference assumptions, that is,
\label{consistency_multi}
$
     Y^{\text{obs}}_{i,t} = Y_{i,t}(w^{\text{obs}}_{1:N,1:T}) = Y_{i,t}(w^{\text{obs}}_{1:N,1:t}) =  Y_{i,t}(w^{\text{obs}}_{i,1:t}) \ \forall \ i = 1,...,N; t = 1,...,T.
$
 \end{assumption}
Assumption~\ref{consistency_multi} generalizes Assumption~\ref{consistency} to the context of multi-sites. The non-anticipating and consistency assumptions are the same as in Assumption~\ref{consistency}. The non-interference assumption, which holds trivially for a single time series, requires that the potential outcomes for one site are only affected by their own treatment path, and not by spillover effects across sites.
In our motivating example, since each county often covers relatively large areas, especially in rural areas, and our study populations are older adults who are less to commute long-distance daily, we expect spillover effects across county boundaries to be small compared to the causal effects of heat alerts within each given county. However, we can not rule out the possibility of the daily movement of people from one county to neighboring counties or that people may be exposed to heat alerts from neighboring counties, which may lead to some spillover effects.

\begin{assumption}[Multi-site Time Series Unconfoundedness]
\label{unconfounded_multi}
The assignment mechanism is unconfounded if for all $W_{i,1:T} \in \mathcal{W} = \{0,1\}^T$, and $\boldsymbol{\mathcal{F}}_t$,
$
W_{i,t} \ \perp \ Y_{i,s}(w_{i,1:s}) \ \mid \ \boldsymbol{\mathcal{F}}_t \ \forall \ i = 1,...,N; t = 1,...,T \ \text{and} \ t \leq s \leq T.
$
\end{assumption}
Assumption~\ref{unconfounded_multi} is similar to Assumption~\ref{unconfounded}, which states the treatment path for each site depends on the past information only. This does not rule out the possibility that the probability of receiving the treatment on day $t$ in one site could depend on covariates observed in another site prior to time $t$. There is no need to assume that the potential outcome in site $i$, $Y_{i,1:t}(w_{i,1:t})$, is independent from the potential outcome in a different site $j$, $Y_{j,1:t}(w_{j,1:t})$, for $i\neq j$. In our motivation example, we adjusted for time-varying confounders identified by \cite{weinberger2021heat}. Collaborating with environmental health experts to collect additional pre-exposure covariates that are associated with both exposures and outcomes and adjust for those covariates may further reduce the confounding bias.

\begin{assumption}[Spatial Random-effects across Multi-sites]
\label{random_multi}
Each site was given the same intervention path $\delta_{1:T}$, and the duration-$t_0$ causal estimands $\bar{\tau}_i(\delta_{(t-t_0):t})$ for site $i$ follows $
      {\bar{\tau}}_i (\delta_{(t-t_0):t,T}) =  \bar{\tau}^{N} (\delta_{(t-t_0):t,T}) + u_i + \epsilon_i,
$
where $\bar{\tau}^{N} (\delta_{(t-t_0):t,T})$ is the weighted-average pooled causal estimand across $N$ sites, $\forall \ t = 1,...,T; t_0 = 1,...,t$, $u_i$ is a random effect to allow for heterogeneity and spatial correlation in the causal effects across site $i = 1,...,N$, and $\epsilon_i$ is sampling variability for site $i$ with known within-site variance, $V_i$.
\end{assumption}
Assumption~\ref{random_multi} has been used in meta-analysis allowing both effect heterogeneity and spatial correlation \citep{dersimonian2015meta,maire2019poleward}.
This assumption consists of two components:
1) we assume the causal estimands at each site are defined by the same ItvPS intervention $\delta_{1:T}$, and thus combining results in a meta-analysis is practically meaningful (e.g., one usually chooses to combine clinical studies of the same drug, rather than completely different drugs, in one meta-analysis); 2) among the well-defined causal estimands $\bar{\tau}_i(\delta_{(t-t_0):t})$, we further assume effects across multi-sites follow a spatial random-effects model. Such a spatial random-effects assumption is previously stated in random-effect meta-analysis literature \citep{maire2019poleward} and was used in multi-site environmental epidemiology studies \citep{bell2004time}. 

We define a weighted average causal estimand across $N$ sites, as the estimand of interest:
\begin{align*}
    \bar{\tau}^N(\delta_{(t-t_0):t,T}) &= \sum^N_{i = 1} c_i \bar{\tau}_i (\delta_{(t-t_0):t,T}), \ \text{where} \ \sum^N_{i = 1} c_i = 1 \ \text{and} \ \{c_i,i =1,2,...,N\} \ \text{are fixed.}
\end{align*}
Such a weighted causal estimand provides an overall summary of the causal effect drawn from a super-population, whereas the observations from each site are treated as a random sample from this super-population \citep{dersimonian2015meta}.
The remaining question is how to choose suitable weights $c_i$ for each site $i$. To answer this question, we propose a random-effects meta-analysis model. 

\subsection{Estimation and Inference}
\label{inf_multi}
The random-effects allow different sites to have heterogeneous causal effects that are sampled from a distribution characterizing the overall causal effect. The primary purpose is to make inferences about the pooled causal effect and provide a quantitative measure of how the causal effects differ across the sites.
In the meta-analysis, to obtain a weighted average causal estimand that better characterizes the pooled causal effect synthesized from multiple sites, we  assign more weight to sites that yield a more precise estimate of the overall causal effect. Random-effect models use an inverse variance scheme to assign weights to each site based on inverse proportions of the total variance from each site.
Specifically, there are two sources of variance under a random-effects model \citep{borenstein2010basic}. First, the observed causal effect $\hat{\bar{\tau}}_{i} (\delta_{(t-t_0):t,T})$ for any time series in one site differs from that time series’s true causal effect because of within-site variance, $V_i$. Second, the true causal effect for each  time series differs from the overall causal effect because of between-site variance, $\Delta^2$.  
\cite{dersimonian1986meta} are among the first to propose the following method to estimate the between-site variance, $\Delta^2$: 
\begin{enumerate}
    \item[1)] Obtain a common effect estimator under a fixed-effect (FE) meta-analysis model,
    $
        \hat{\bar{\tau}}_{FE} = \frac{\sum_{i=1}^N \hat{\bar{\tau}}_{i} (\delta_{(t-t_0):t,T})/V_i}{\sum_{i=1}^N 1/V_i}.
    $
    \item[2)] Based on $\hat{\bar{\tau}}_{FE}$, calculate the Cochran's Q‐statistic,
    $
        Q = \sum^N_{i=1} \frac{[\hat{\bar{\tau}}_{i} (\delta_{(t-t_0):t,T})- \hat{\bar{\tau}}_{FE}]^2}{V_i}.
    $
    \item[3)] Obtain the estimator for the between-site variance $\Delta^2$ as
    $
     \hat{\Delta}^2_{DL} = \frac{Q - (N-1)}{\sum_{i=1}^N 1/V_i - \frac{\sum_{i=1}^N 1/V^2_i}{\sum_{i=1}^N 1/V_i}}.
    $
\end{enumerate}
The weight assigned to each site under the inverse variance scheme is
$
    c_i = \frac{1}{V_i + \Delta^2}, \ i =1,2,...,N,
$
where the within-site variance, $V_i$, is unique to each site, whereas the between-site variance $\Delta^2$ is a quantity that is common across sites. Various methods to estimate $\Delta^2$ are summarized in \cite{borenstein2010basic}. To allow spatial correlation in the causal effects across sites, we additionally use a spatial random-effects meta-analysis model proposed by \cite{maire2019poleward}. For this model, we first identify the spatial coordinates (longitude and latitude) for the centroids of each county. We then specify a Gaussian spatial correlation structure based on the relative Euclidean distances computed by the longitude and latitude, $d$. Finally, we fit the following spatial random-effects meta-analysis model using \textbf{metafor} R package \citep{viechtbauer2010conducting}:
    $
      {\bar{\tau}}_i (\delta_{(t-t_0):t,T}) =  \bar{\tau}^{N} (\delta_{(t-t_0):t,T}) + u_i + \epsilon_i,
    $
    where ${\bar{\tau}}_i(\delta_{(t-t_0):t})$ is the duration-$t_0$ causal effects for site $i$, $\bar{\tau}^{N} (\delta_{(t-t_0):t,T})$ is the weighted-average pooled causal effect across $N$ sites. 
    The random effects $u_1,\ldots,u_N$ were assumed to follow a multivariate normal distribution with a mean of zero and a variance-covariance matrix $
        Cov(u_i, u_j) = \Delta^2 \times \exp(-d_{i,j}^2/\rho^2),
    $
    where $\Delta^2$ denotes the between-site variance, $d_{i,j}$ denotes the Euclidean distance between two spatial points $i$ and $j$. $\rho$ is the spatial correlation parameter for the Gaussian correlation structure, which can be estimated by restricted maximum likelihood (REML).
\vspace{-0.5cm}
\section{Simulation Study}
\label{sec:sim}
We study finite-sample properties of the proposed estimators via simulation studies on a single time series, in which we vary: a) the length of the time series $T$; b) the duration of interventions $t_0$; and c) the assignment mechanisms of the treatment $p_t(W_t = 1 \mid \mathcal{F}_t)$. We conduct a comparison comparing the IPW estimator to the nonparametric influence-function-based estimator initially proposed by \cite{kennedy2019nonparametric}, modified to the time series setting.
To reflect the nature of treatment assignment in our motivating example, we generate the treatment $W_t$ under a nearly non-overlapping setting.
In particular, we consider a time series with length $T$. For each $t=1,\ldots,T$,
$\textbf{C}_t = (C_{1,t}, C_{2,t}, C_{3,t}, C_{4,t}, C_{5,t}) \sim N(\textbf{0}, \textbf{I}_5); \
    p_t(W_t = 1 \mid \mathcal{F}_t) =  \text{expit}\{10 \times (\sum^5_{j=1} {C}_{j,t}/5 - W_{t-1} + 0.5)\}; \
    Y_t \mid W_t, W_{t-1},\textbf{C}_t \sim N(3 \times W_t + W_{t-1} + \sum^5_{j=1} {C}_{j,t}/5,1).
$
At time $t$, the random assignment mechanism of the treatment $W_t$  depends on both $\textbf{C}_t$, and the treatment at time $t-1$, $W_{t-1}$.  Also the outcome $Y_t$ depends on the treatments both at time $t$ and $t-1$ (duration-$1$ effect). The data generating mechanism described above mimics a nearly non-overlap setting (see Figure S.1 of the Supplementary Materials, showing the distributions of the time-varying propensity scores have little overlap across treated vs. untreated units).
The main quantity of interest is the duration-$t_0$ causal estimand on the observed treatment path. We assess the performance of each estimator by calculating the integrated bias and root mean squared error (RMSE) defined as;
$
\widehat{\text{Integrated Bias}} = \frac{1}{J} \sum^J_{j=1} \Big|\frac{1}{K} \sum^K_{k=1} \big\{ \hat{\bar{\tau}}^k(\delta_{t-t_0:t,T,j}) - \bar{\tau}^k(\delta_{t-t_0:t,T,j}) \big\}\Big|, 
\widehat{\text{RMSE}} = \frac{\sqrt{N}}{J} \sum^J_{j=1} \Big[\frac{1}{K} \sum^K_{k=1} \big\{\hat{\bar{\tau}}^k(\delta_{t-t_0:t,T,j}) - \bar{\tau}^k(\delta_{t-t_0:t,T,j})\big\}^2\Big]^{1/2},
$
where $\bar{\tau}^k(\delta_{t-t_0:t,T,j}) := \frac{1}{T-t_0} \sum^T_{t=t_0+1} \bar{\tau}^k(\delta_{t-t_0:t,j})$ is the true temporal average causal quantity, and $\hat{\bar{\tau}}^k(\delta_j)$ is its estimator based on the estimated time-varying propensity score, in which the superscript $k$ indicates the simulation replicate. We assess the estimation performances at $J = 50$ values of $\delta_j$ equally spaced between $0.1$ to $10$ evaluated on $K = 500$ simulation replicates. We also assess the average coverage of our proposed point-wise Wald 95\% CIs, and the coverage of time-uniform 95\% CSs.

We vary the following combinations of $(T,t_0)$, $T = (200,1000,5000)$, and $t_0 = (1,4,9)$. We apply both parametric and nonparametric models to estimate the time-varying propensity scores, following the recommendation by \cite{bonvini2021incremental}, 1) logistic regression; 2) Super Learner \citep{van2007super}, which combines generalized additive models, multivariate adaptive regression splines, support vector machines, and random forests, along with parametric generalized linear models (with and without interactions, and with terms selected stepwise via AIC), consistent with \cite{kennedy2019nonparametric} (implemented by the \textbf{SuperLearner} R package). We assume that the form of the time-varying propensity score model is correctly specified as $W_t \mid \textbf{C}_t, W_{t-1}$ with a \textbf{logit} link. Note that the IPW estimator relying on the Super Learner for propensity score estimation does not maintain the same asymptotic properties as those using a correctly specified parametric logistic regression model. We conduct additional simulations in the Supplementary Materials Section S.3 when the form of the time-varying propensity score model is misspecified.

Table~\ref{sim:results} shows the integrated bias and RMSE of the IPW and influence-function-based estimator, along with the average coverage rate of their corresponding Wald 95\% CIs. Under the same duration $t_0$, we observed that the integrated bias and RMSE of the estimator generally decrease when the length of the time series, $T$, increases, regardless of how the time-varying propensity score is estimated, either by logistic regression or Super Learner.
We also observed a decreased performance of the estimator as the duration, $t_0$, increases. This finding is not surprising given that  the estimated weights in the proposed weighting estimator depend on the product of $t_0+1$ terms of estimated ItvPS. We expect that the proposed estimator will likely be unstable and thus the absolute bias and RMSE will be larger when $t_0$ increases. In practice the exact duration $t_0$ is unknown and researchers need to specify $t_0$ based on their prior domain knowledge. 
The simulation results suggest that the choice of $t_0$ should be parsimonious, i.e., one should choose a duration $t_0$ that can capture the data complexity yet is as small as possible. We also found that using the Super Learner model to estimate the time-varying propensity score leads to improved performances compared to the logistic regression model even if the propensity score model is correctly specified with a \textbf{logit} link. This finding suggests that the Super Learner, as an ensemble of flexible parametric/nonparametric models, has good finite-sample performances in various settings. We did not find clear advantages of the influence-function-based estimator in finite-sample performances compared to the IPW estimator, under this simulation setting. In particular, we found the IPW estimator achieves smaller integrated bias and RMSE when the sample size is relatively large ($T=1000,5000$), compared to the influence-function-based estimator. Additional simulations in the Supplementary Materials Section S.3 indicate promising performances of the Super Learner when the underlying data generating mechanism is unknown and potentially misspecified.  

\begin{figure}[ht]
\centering
\includegraphics[width=1\linewidth]{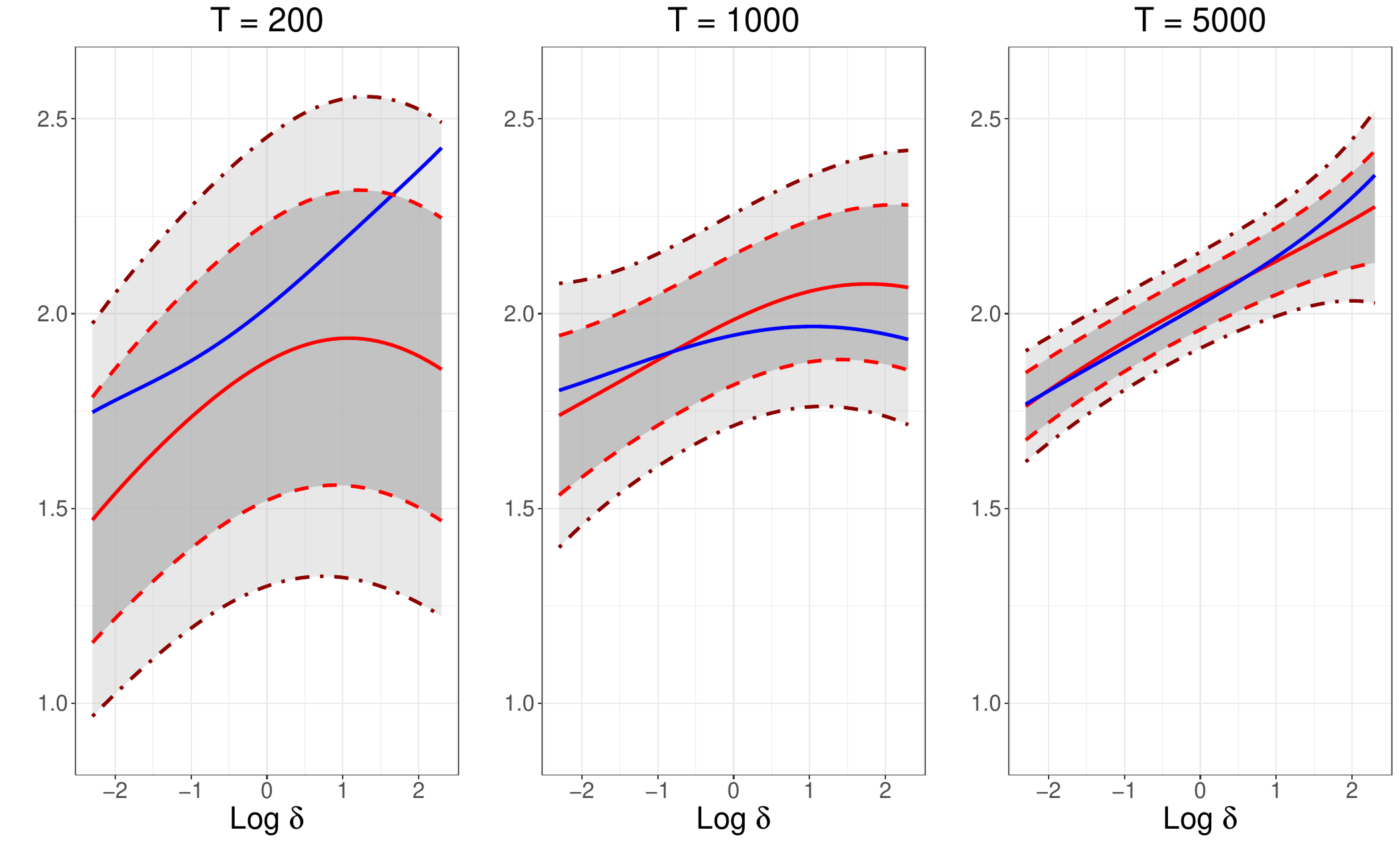}
\caption{The duration-$1$ causal effect curves obtained by the IPW estimation when the probability of treatment assignments was multiplied by an odds ratio $\delta_t \in [0.1,10]$. The red solid line represents the estimated causal effect curve along with the point-wise Wald 95\% CIs (red dashed line) and the time-uniform 95\% CSs (dark red dot-dash line). The blue solid line represents the true causal effect curve. The left, middle and right panels reflect the simulation scenarios with $T = 200, 1000, 5000$. The propensity scores were estimated by a Super Learner algorithm including generalized additive models, multivariate adaptive regression splines, support vector machines, and random forests, along with parametric generalized linear models (with and without interactions and with terms selected stepwise via AIC), consistent with \cite{kennedy2019nonparametric}.
}
\label{ts_ips}
\end{figure}

We also found that the coverage rates of the Wald 95\% CIs for the IPW estimator were near or above the nominal level (95\%) when $T$ is relatively large, as shown in Table~\ref{sim:results}. The conservative performance is likely due to the fact that we used an upper bound for the variance estimates when constructing the CIs. The coverage rates of the Wald 95\% CIs for the influence-function-based estimator were more frequently below the nominal level. The coverage rates decrease when $t_0$ increases, which is consistent with the coverage results shown in \cite{papadogeorgou2020causal}. Figure~\ref{ts_ips} visualizes the curve of duration-$1$ causal effects when the probability of the treatment assignment are multiplied by odds ratios $\delta_t \in [0.1,10]$. The red solid line represents the estimated causal effects along with point-wise Wald 95\% CIs (red dashed line). We found that the bias between the estimated curve and the true curve reduces when $T$ increases. The point-wise Wald 95\%  CIs  capture the true curve in all three scenarios; $T = 200, 1000, 5000$.  The time-uniform CSs generally perform unstably when the sample sizes are small ($T = 200$), yet perform conservatively when the sample sizes are relatively large ($T = 1000,5000$).

We design our simulation studies to reflect scenarios of nearly non-overlap between treated and untreated units observed in our data application. Our simulation results show that the proposed estimators perform well in terms of bias and root mean squared error even when the overlap assumption is nearly violated. 
\vspace{-0.5cm}
\section{Application}
\label{sec:app}
We analyze the multi-site time series data described in Section~\ref{motivate_data}.
First, we apply the proposed methods to estimate the causal estimands $\bar{\tau}_i(\delta_{(t-t_0):t,T})$.
We assume $\delta_t$ is the same for every day $t$ during the warm months (April-October) of 2006-2016. In the time-varying propensity score model, we included the following observed covariates: daily maximum heat index, lag-1 daily maximum heat index, lag-2 daily maximum heat index, moving average heat index during the current warm season, lag-1 day heat alert, lag-2 day heat alert, the running total number (the summation of the sequence of numbers updated daily) of heat alerts that have been issued during the current warm season, the moving average number of deaths/hospitalizations during the current warm season, day of the week, and federal holidays. Note, as described in Section~\ref{motivate_data}, the inclusion of past treatments (lagged heat alerts) and past outcomes (historical deaths/hospitalizations) is allowed in the time-varying propensity score model.  The time-varying propensity score was estimated by a Super Learner using the same combination of algorithms described in Section~\ref{sec:sim}.

For each county $i$, we estimate the daily numbers of all-cause deaths and cause-specific hospitalizations for five heat-related diseases (heat stroke, urinary tract infections, septicemia, renal failure, fluid and electrolyte disorders) under several ItvPS intervention scenarios ranging from  $\delta_t = 1$ to $\delta_t = 10$. We consider the causal estimands with duration $t_0 = 2$, consistent with \cite{weinberger2018effectiveness}. We define the county-specific causal effect curve for county $i$ as $\bar{\tau}_i(\delta_{(t-2):t,T}) - \bar{\tau}_i(1,1,1)$, where $\bar{\tau}_i(\delta_{(t-2):t,T})$  denotes the estimated daily average number of deaths or hospitalizations under various ItvPS interventions ($\delta_t \in [1,10]$) and $\bar{\tau}_i(1,1,1)$ denotes the quantity corresponding to the daily average number of deaths or hospitalizations observed \textit{factually} (as if there was no change in the treatment assignments).
After obtaining all $N = 2,837$ county-specific causal effect curves, we utilize the spatial random-effects meta-analysis approach proposed in Section~\ref{sec:meta} to pool the estimated county-specific causal effect curves across multiple counties, and obtain the estimated pooled causal effect curve, $\sum^N_{i=1} c_i [\bar{\tau}_i(\delta_{(t-2):t,T}) - \bar{\tau}_i(1,1,1)]$, where the weights $c_i, i = 1,2,\ldots, N$ are obtained by the spatial random-effects meta-analysis model.

\begin{figure}[ht]
\noindent\begin{minipage}{0.5\columnwidth}
\includegraphics[width=1\linewidth]{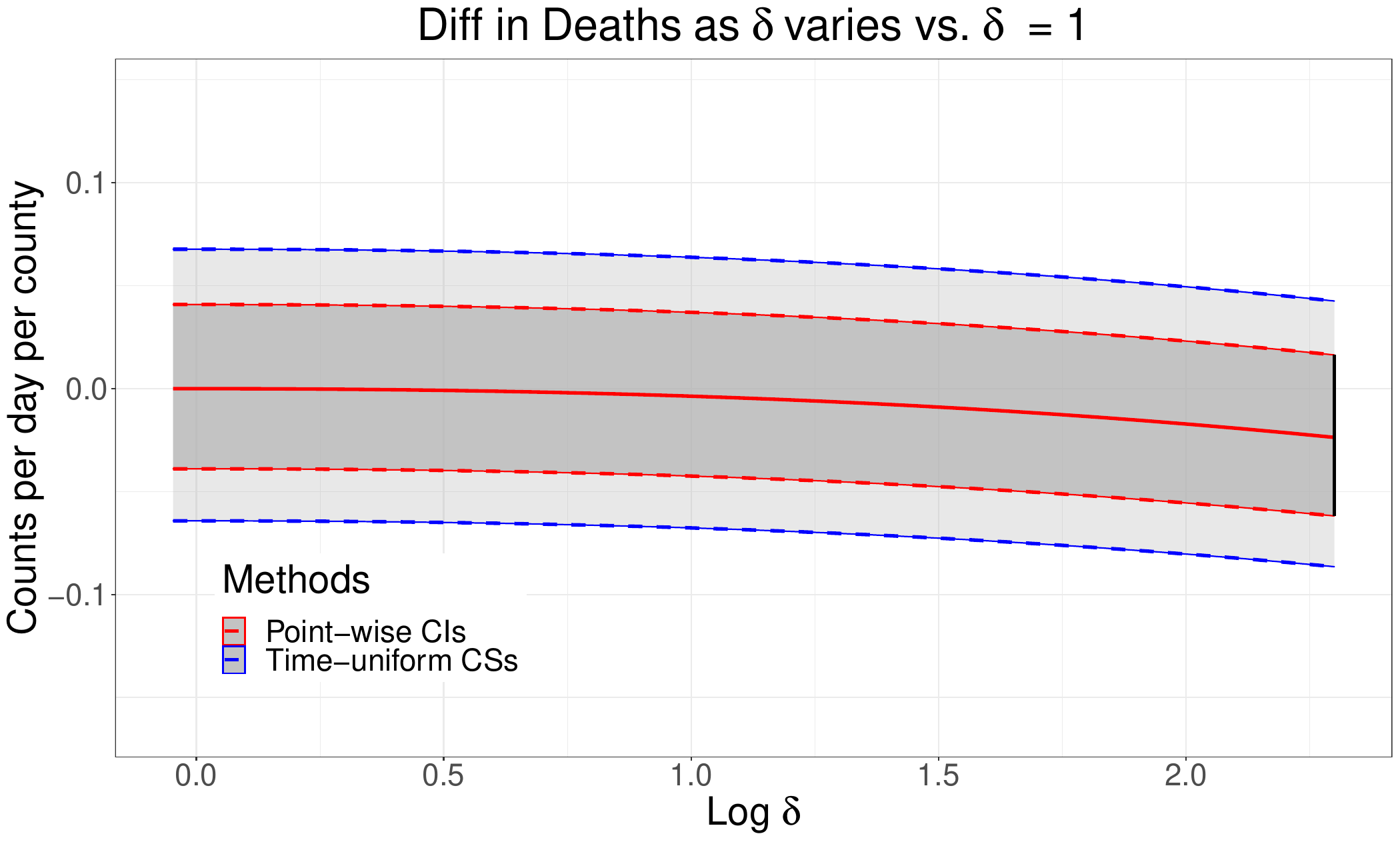}
\end{minipage}
\begin{minipage}{0.5\columnwidth}
\includegraphics[width=1\linewidth]{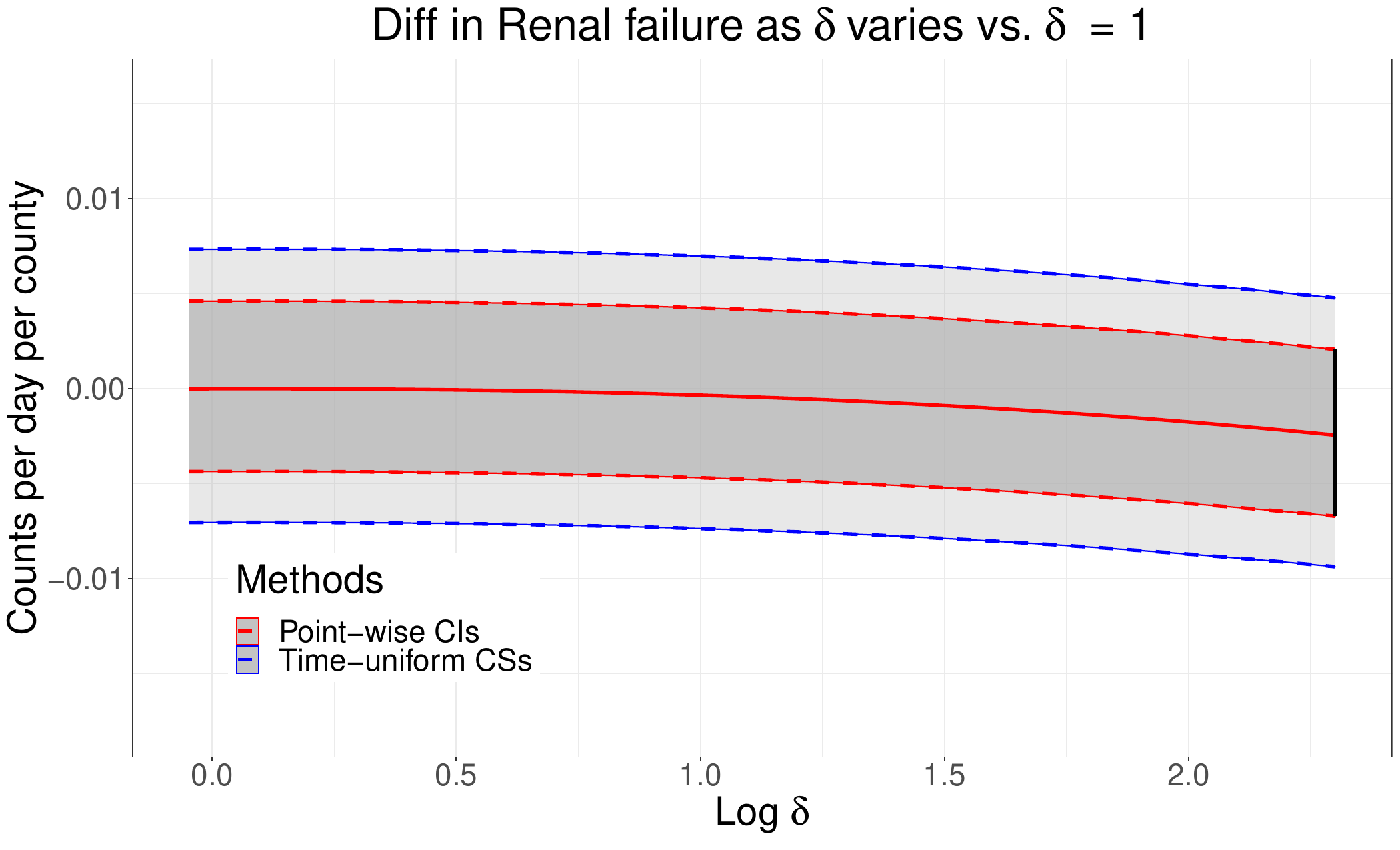}
\end{minipage}
\begin{minipage}{0.5\columnwidth}
\includegraphics[width=1\linewidth]{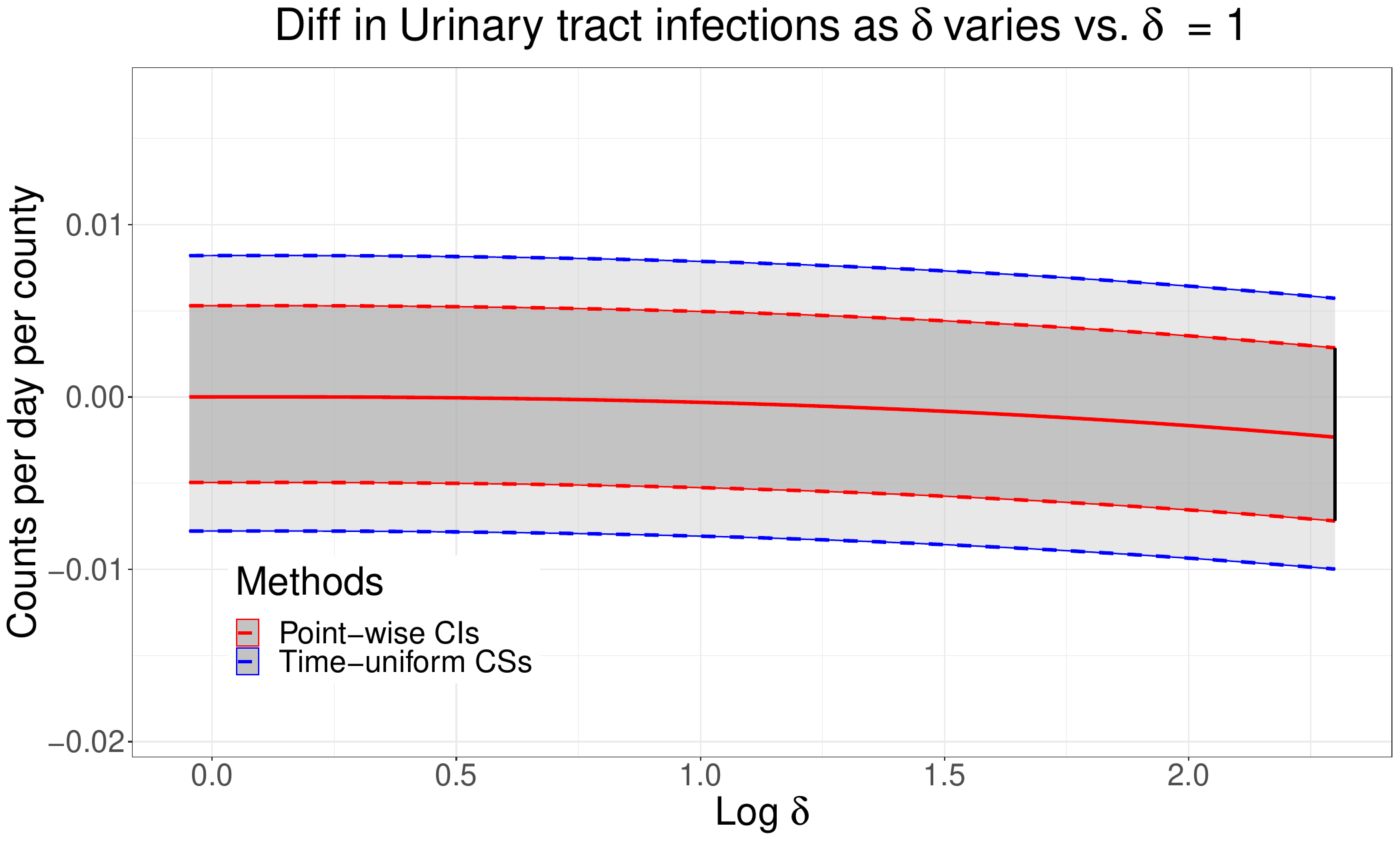}
\end{minipage}
\begin{minipage}{0.5\columnwidth}
\includegraphics[width=1\linewidth]{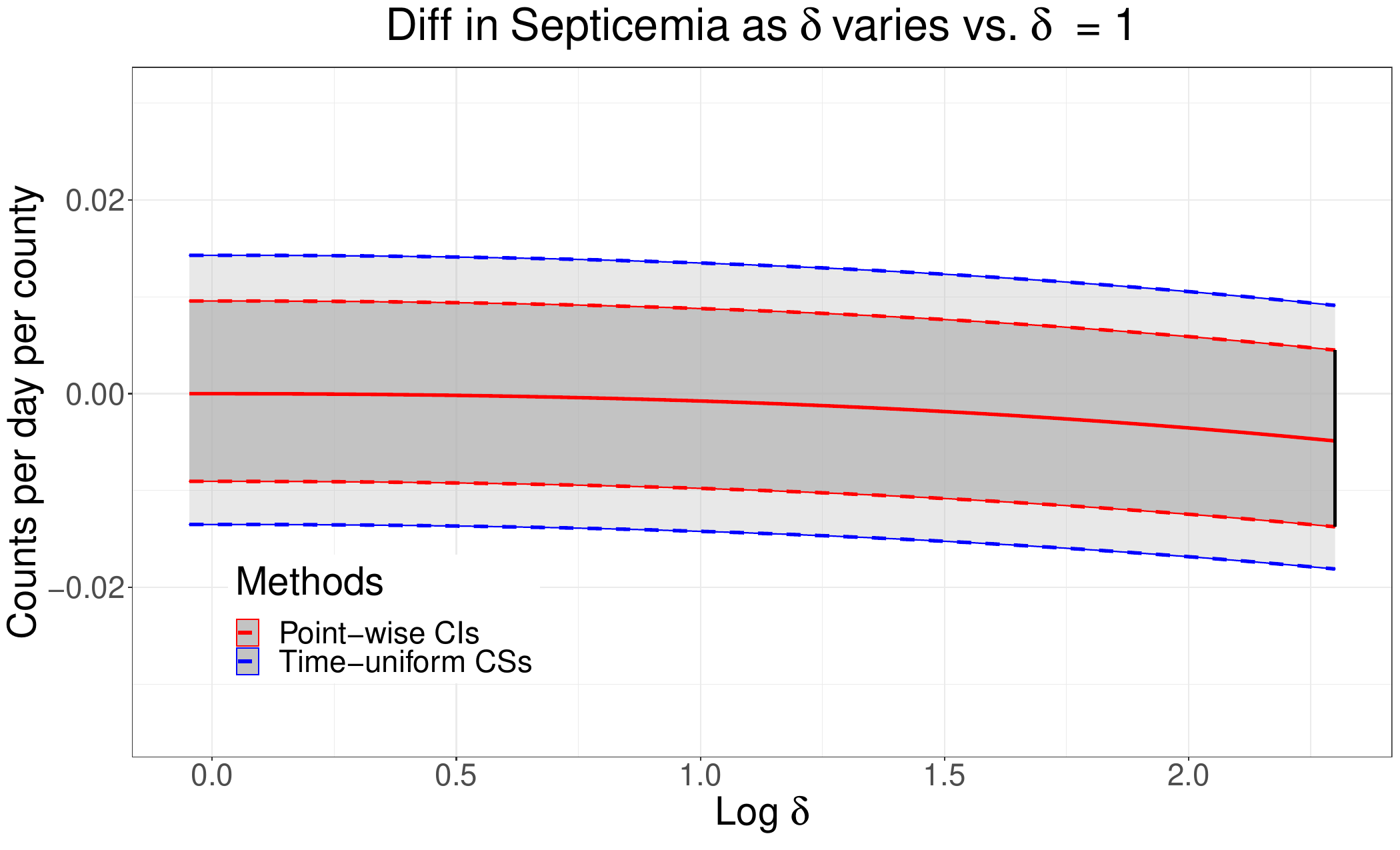}
\end{minipage}
\begin{minipage}{0.5\columnwidth}
\includegraphics[width=1\linewidth]{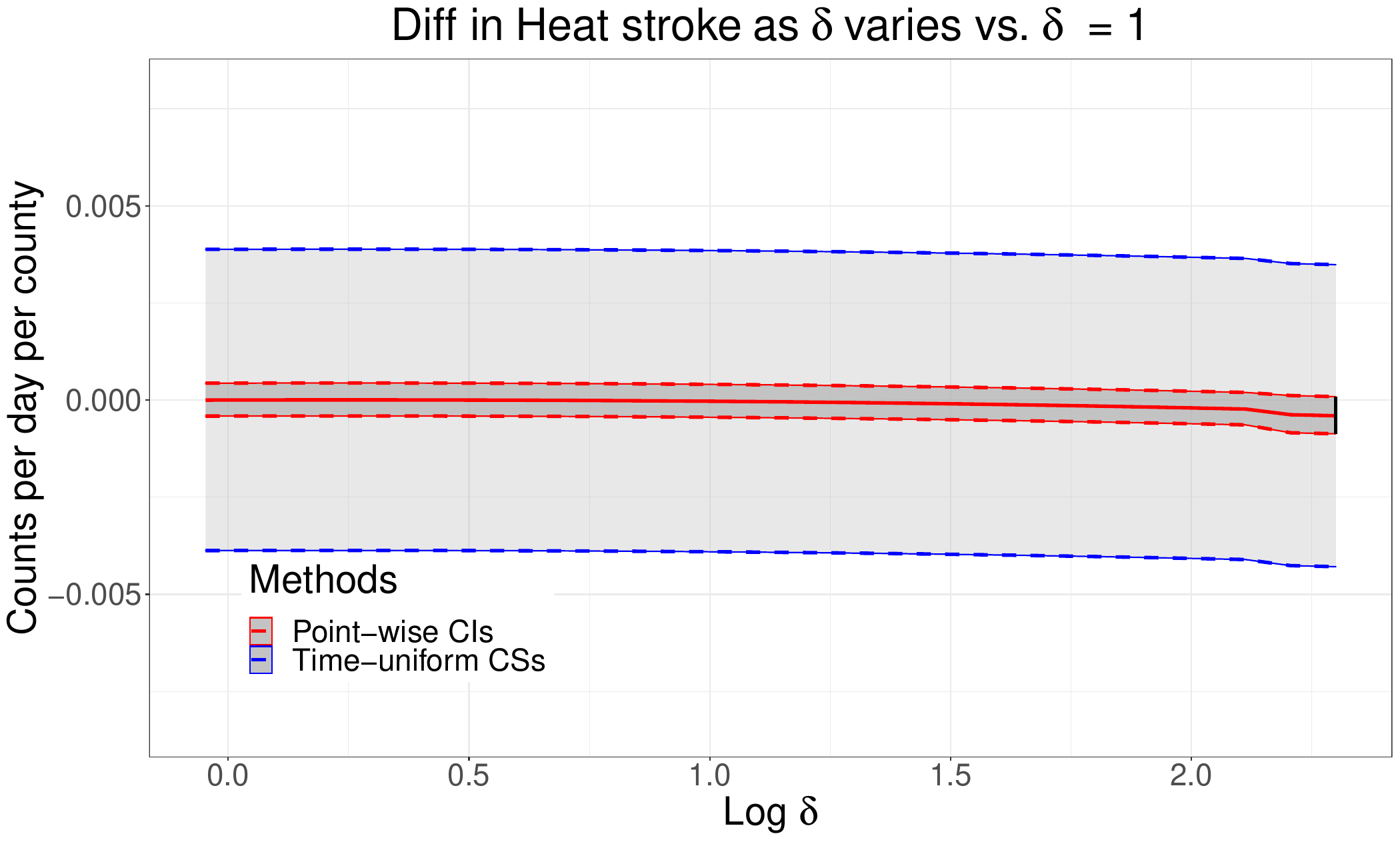}
\end{minipage}
\begin{minipage}{0.5\columnwidth}
\includegraphics[width=1\linewidth]{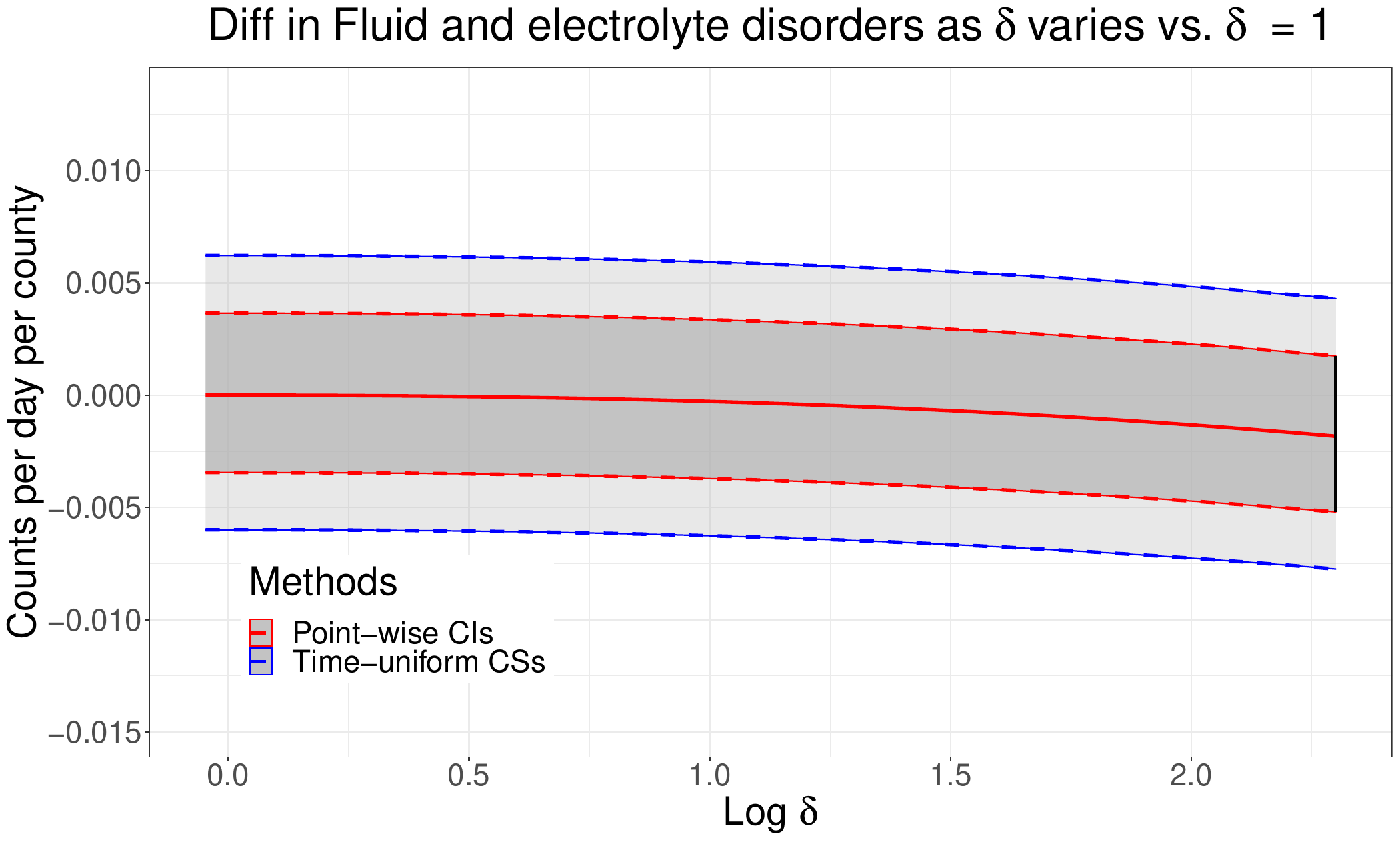}
\end{minipage}
\caption{\label{hosp_result} The estimated pooled causal effect curves of average all-cause deaths and cause-specific hospitalizations for five heat-related diseases per day per county among $2,837$ counties, assuming the duration of interventions was $2$ days. The curves represent the differences in deaths and hospitalizations comparing the counterfactual situations where the probability of issuing heat alerts was multiplied by an odds ratio $\delta_t \in [1,10]$ to the factual situations where the probability of issuing heat alerts remains unchanged ($\delta_t = 1$). The dashed red lines represent the corresponding point-wise Wald 95\% CIs of the differences. The dashed blue lines represent the time-uniform 95\% CSs. 
}
\end{figure}

Figure~\ref{hosp_result} shows the estimated pooled causal effects for the daily average number of all-cause deaths and cause-specific hospitalizations for five heat-related diseases per county among $2,837$ counties across the warm months of 2006-2016. The curves represent the differences in daily average numbers of deaths and hospitalizations averaged across $2,837$ counties  comparing the counterfactual scenarios where the probability of issuing heat alerts was multiplied by an odds ratio $\delta_t \in [1,10]$ to the factual scenario, where the probability of issuing heat alerts remains unchanged ($\delta_t = 1$).
The dashed red lines represent the corresponding point-wise Wald 95\% CIs of the differences. The dashed blue lines represent the time-uniform 95\% CSs. The black vertical lines represent the average number of deaths and hospitalizations that could be avoided per day per county and their corresponding CIs if we had increased the probability of issuing heat alerts by the maximum odds considered,  $\delta_t = 10$. We found, in general, consistent downward patterns among the pooled causal effect curves for each health outcome as the $\log(\delta_t)$ increases above 0, indicating slight reductions in average all-cause deaths and cause-specific hospitalizations for five heat-related diseases among Medicare enrollees as $\log(\delta_t)$ increases. The CIs contain $0$ throughout the range of $\delta_t \in [1,10]$.

Quantitatively, we found that if we had increased the probability of issuing heat alerts by an odds ratio $\delta_t = 10$, on average $0.15 \ (95\%\ \text{CI:} -0.01 \ \text{to} \ 0.32)$ deaths could be averted per day per county (see vertical black line in Figure~\ref{hosp_result}\textbf{a}).
Based on these estimates, solely among extremely hot days (i.e., the top 5\% hottest days when the heat alerts are more likely to be issued), we estimated $4,653$ avoidable deaths $(95\%\ \text{CI:} -415 \ \text{to} \ 9,576)$ across all $2,837$ counties in one warm season. Similarly, if we increase the probability of issuing heat alerts by an odds ratio $\delta_t = 10$ the number of hospitalizations averted are $42 \ (95\%\ \text{CI:} -7 \ \text{to} \ 91)$ for heat stroke; $634  \ (95\%\ \text{CI:} -81 \ \text{to} \ 1,327)$ for urinary tract infections; $1,157  \ (95\%\ \text{CI:} -134 \ \text{to} \ 2,409)$ for septicemia; $585 \ (95\%\ \text{CI:} -18 \ \text{to} \ 1,169)$ for renal failure; $405 \ (95\%\ \text{CI:} -43 \ \text{to} \ 840)$ for fluid and electrolyte disorders among extremely hot days in one warm season. However, given the wide CIs for the results of all outcomes, we did not find any statistically significant causal effects of increasing the probability of issuing heat alerts on health outcomes comparing $\delta_t = 10$ (increased odds) vs. $\delta_t = 1$ (unchanged). We also found there is large between-county heterogeneity in the random-effects meta-analysis model, $(\text{P-value of heterogeneity test} < 0.001)$, for all outcomes. Overall, our findings suggest weak evidence that increasing the probability of issuing heat alerts may bring health benefits to the U.S. Medicare population.

In the Supplementary Materials Section S.4, we conducted additional county-specific analyses that include time series data from three selected counties: Santa Clara, CA; Maricopa, AZ; and New York, NY, separately. Notably, we found that in Maricopa, AZ, increasing the probability of issuing heat alerts could significantly reduce all-cause deaths, whereas less dramatic but still statistically significant results were found in Santa Clara, CA, and New York, NY (see Figure S.2).
As a sensitivity analysis, we analyzed a subset of $550$ counties with population sizes $> 100,000$ (``populous counties").  These counties included more than $70\%$ of all-cause deaths and cause-specific hospitalizations observed among the $2,837$ counties. 
As shown in Figure S.3, the shapes of the causal effect curves based on the subset of $550$ populous counties are very similar to the estimated curves based on data from all $2,837$ counties. We found non-significant results for all outcomes. 
The random-effects meta-analysis model still indicates considerable between-county heterogeneity $(\text{P-value of heterogeneity test} < 0.001)$ for  all outcomes.
Additional analysis was conducted in which the duration of interventions was varied. We also estimated pooled causal effect curves assuming the duration of interventions was $0$ and $5$ days respectively. As shown in Figure S.4-5, the causal effect of increasing the frequency of issuing heat alerts may have more pronounced effects on health outcomes when considering a longer duration of interventions. Importantly, we observed statistically significant results of the causal effects of increasing the probability of issuing heat alerts on health outcomes comparing $\delta_t = 10$ (increased odds) vs. $\delta_t = 1$ (unchanged) when considering cumulative effects up to duration $t_0 = 5$, i.e., on average $0.15 \ (95\%\ \text{CI:} \ 0.06 \ \text{to} \ 0.24)$ avoidable deaths per day per county. We also found the widths of CIs for the causal effect curves of duration-$0$, duration-$2$ and duration-$5$ estimands did not change notably as $t_0$ varies.
\vspace{-0.5cm}

\section{Discussion}
\label{sec:conc}
We have developed a novel causal inference framework for multi-site time series data that relies on stochastic interventions. Under this framework, we introduced a class of causal estimands and their unbiased estimators  with theoretical justification. In the context of multi-site  time series data, we link our ItvPS intervention framework to the spatial random-effects meta-analysis. The ultimate goal is to obtain a pooled causal estimator that summarizes the potentially heterogeneous causal effects from multi-site time series studies.

Our approach was motivated by our data application, where the goal was to assess  whether or not increasing the probability of issuing heat alerts reduces  morbidity and mortality.
More specifically, we estimated the causal effects of increasing the probability of issuing  heat alerts on  all-cause deaths  and cause-specific hospitalizations for five heat-related diseases.  We found some evidence of reductions in morbidity and mortality  with large statistical uncertainty. 
Importantly, the random-effects meta-analysis model indicated large heterogeneity of the causal effects across different counties. 

One key distinction of our approach compared to more traditional statistical methods in environmental epidemiology is that most time series studies in this context use matched case-crossover or difference-in-difference designs, where the focus is often the (average) causal contrast between treated days (i.e.,  days with heat alerts) vs. untreated days (i.e.,  days without heat alerts), and the number of heat alert days is then fixed for a given study period in one chosen site by the study design \citep{chau2009hot,weinberger2018effectiveness,weinberger2021heat}. In contrast, our analysis focuses on the causal effect of a stochastic intervention, estimating the causal effect as the probability of issuing heat alert changes, and as such, the counterfactual heat alert days are not fixed. Therefore our results are not directly comparable to previously published studies \citep{weinberger2021heat}.  

While our proposed novel causal inference approach  provides a new powerful tool in policy evaluations, some methodology considerations are needed when applying this approach to future studies.
First, we expect causal effects defined by stochastic interventions to answer a different policy question than that of a deterministic intervention \citep{kennedy2019nonparametric}. The proposed causal estimand is an intuitive quantity answering the following causal question ``how many adverse health outcomes could have been averted if we change the frequency of heat alerts?" which was the focus of our work. However, other practitioners' interests may be more aligned with estimating causal effects of deterministic interventions, for instance, ``how many adverse health outcomes could have been avoided if the temperature was, on average, one degree lower during extremely hot days \citep{weinberger2020estimating}?" In that case, causal inference methods based on deterministic interventions \citep{bojinov2019time,bojinov2020panel,rambachan2021common} may be used. 
Second, the proposed weighting estimator maintains ideal theoretical properties only when the time-varying propensity scores can be modeled using correctly specific parametric models. While an influence-function-based estimator may be compatible with the nonparametric propensity score estimation, its theoretical properties  under the time series setting require further investigation. As part of future work, we plan to conduct theoretical analysis of doubly-robust targeted estimation of an ItvPS intervention.
Third, the random-effects meta-analysis creates a weighted average causal estimand and its corresponding pooled estimator, in which the weights are calculated by combining  between-site and within-site variance. For this reason, the weighted average causal quantity is defined on a hypothetical weighted population. \cite{dahabreh2020toward} have criticized that standard meta-analyses may produce results that do not belong to a clear target population when each site represents a different population and the treatment effect varies across these populations. As part of future work, we plan to develop approaches that allow inferences to be transported from multi-site time series to a clearly specified target population \citep{dahabreh2020toward}. 
Fourth, the non-interference assumption may not hold in many climate and health studies, such as this one, which rely on spatial-temporal data. For instance, the NWS-issued heat alerts in one county may impact people in adjacent counties. As part of future work, we hope to extend the time series intervention path to a multivariate intervention path defined by random matrices \citep{papadogeorgou2020causal}, to potentially overcome the violation of non-interference assumption, and identify direct and spillover effects under this stochastic intervention framework. Fifth, the stable estimation of time-varying propensity score based on time series observational data is challenging since the time-varying confounder sets are potentially high-dimensional. While in our data application we used the state-of-the-art Super Learner (an ensemble of flexible parametric/nonparametric models) to estimate time-varying propensity scores, we had to limit the size of confounder sets to avoid unstable estimates due to the curse of dimensionality. Also, the proposed weighting estimator that relies on products of the estimated ItvPS may be unstable when accounting for the longer duration of interventions.
We plan to generalize covariate balance methods to improve high-dimensional propensity score estimation and stability of the weighting estimators in time series observational studies \citep{athey2018approximate}.

The stochastic intervention framework for time series data introduced in this paper is the first approach that allows the identification and estimation of causal quantities defined by stochastic interventions on multi-site time series. We believe this framework addresses one of the emerging methodological needs in climate and health research, where researchers often collect time series data from multiple geographic locations seeking causal evidence among diverse populations \citep{liu2019ambient,lee2020projections}. Furthermore, we expect the proposed framework can be applied to science and policy-relevant research in political science, economics, and law, where a considerable amount of spatial-temporal data are generated and collected.

%\section*{Supplementary material}
%Supplementary material is available online at \url{http://biostatistics.oxfordjournals.org}.
\vspace{-0.5cm}

\section*{Acknowledgement}
The authors are grateful to Jose R. Zubizarreta, Ambarish Chattopadhyay, Eli Ben-Michael, Kosuke Imai, and Guanbo Wang for helpful discussions.  Funding was provided by the National Institute of Health (NIH) grants R01ES029950, R01AG066793(-02S1), R01ES030616, R01AG060232-03, R01ES028033(-S1), R01MD012769, R01ES026217, 1RF1AG071024, 1RF1AG074372-01A1, P01AG031720, P30ES000002, F32ES027742; Alfred P. Sloan Foundation grant G-2020-13946; grant 216033-Z-19-Z from the Wellcome Trust, Harvard University Climate Change Solutions Fund, and Fernholz Foundation. The funders had no role in considering the study design or in the collection, analysis, interpretation of data, writing of the report, or decision to submit the article for publication. Dr. Wellenius and Dr. Dominici have received consulting income from the Health Effects Institute. Dr. Wellenius recently served as a visiting scientist at Google, LLC. The authors report that they have no conflicts of interest relevant to this work. 
%} \fi

\bibliographystyle{biorefs}
\bibliography{xiaowu}

\newpage
\markboth%
% First field is the short list of authors
{X. Wu and others}
% Second field is the short title of the paper
{Stochastic intervention in time series}

\begin{table}[ht]
\centering
\caption{\label{character} Characteristics for NWS-issued heat alerts, all-cause deaths among Medicare enrollees, and cause-specific hospitalizations for five heat-related diseases among Medicare FFS enrollees across April-October of 2006-2016.}
\begin{tabular}{lcc}
\hline
Variables  &  $2,837$ Counties &  $550$ Populous Counties$^{1}$  \\
\hline

\% Days with Heat Alerts  &  2.52 \%  & 2.22 \%  \\
\# of Deaths &   10,467,201 & 7,653,987 \\
\# of Heat stroke       &  97,399     & 72,649     \\
\# of Urinary tract infections    &  1,424,046  & 1,061,060   \\
\# of Septicemia                  &  2,614,871  & 1,954,011   \\
\# of Renal failure               &  1,207,903  & 894,164    \\
\# of Fluid and electrolyte disorders & 928,270 & 673,007\\
                \hline
\end{tabular}

1. Counties with population $> 100,000$.
\end{table}

\begin{table}[ht]
\centering
\caption{\label{sim:results}  Simulation results for the scenario assuming the treatment assignment mechanism is specified with a \textbf{logit} link. The integrated bias and root mean squared error (RMSE) (multiplied by 10 for easier interpretation) of proposed estimators, the average coverage of proposed confidence intervals (CIs), and the uniform coverage of proposed confidence sequences (CSs) for IPW estimators.}
\begin{tabular}{cccccccccc}
\hline
Propensity Score & & & \multicolumn{3}{c}{IPW} & \multicolumn{2}{c}{influence-function-based} \\
Model & $T$ & $t_0$ & Bias (RMSE) & Coverage (\%) & Uniform Coverage (\%) & Bias (RMSE) & Coverage (\%) \\
\hline
\multirow{9}{2.5cm}{Logistic Regression}  & \multirow{3}{0.8cm}{200}   
  &     1& 1.34 (1.57) & 99.97  & 98.20 & 1.31 (1.57) & 99.84  \\
  &   & 4& 2.95 (3.15) & 97.80  & 98.20   & 2.80 (3.03) & 96.56  \\
  &   & 9& 5.57 (5.73) & 89.58  & 16.40 & 5.18 (5.38) & 84.22 \\
  & \multirow{3}{0.8cm}{1000} 
  &    1   & 0.25 (0.45) & 100.00 & 100.00 & 0.30 (0.48) & 98.06  \\
  &  & 4   & 0.59 (0.74) & 99.98  & 100.00 & 0.78 (0.90) & 79.30 \\
  &  & 9   & 1.06 (1.18) & 96.45  & 100.00 & 1.60 (1.68) & 65.46 \\
  & \multirow{3}{0.8cm}{5000} 
  &    1  & 0.12 (0.20) & 100.00 & 100.00 & 0.16 (0.23) & 99.42  \\
  &  & 4  & 0.30 (0.35) & 100.00 & 100.00 & 0.40 (0.43) & 91.84 \\
  &  & 9  & 0.60 (0.64) & 97.68  & 100.00 & 0.78 (0.81) & 70.41   \\
                \hline
\multirow{9}{2.5cm}{Super Learner}  & \multirow{3}{0.8cm}{200}   
  &     1  & 1.08 (1.33) & 99.64 & 100.00 &  1.26 (1.50) & 99.16  \\
  &   & 4  & 2.24 (2.46) & 95.49 & 49.00 & 2.39 (2.62) & 95.24  \\
  &   & 9  & 4.11 (4.30) & 90.89 & 0.00  & 4.23 (4.43) & 82.65  \\
  & \multirow{3}{0.8cm}{1000} 
  &    1   & 0.04 (0.34) & 100.00 & 100.00 & 0.08 (0.35) & 99.97 \\
  &  & 4   & 0.08 (0.38) & 99.99  & 100.00 & 0.25 (0.46) & 99.73 \\
  &  & 9   & 0.21 (0.49) & 99.41  & 100.00 & 0.40 (0.59) & 98.70 \\
  & \multirow{3}{0.8cm}{5000} 
  &    1  & 0.04 (0.16) & 100.00&  100.00 & 0.04 (0.16) & 100.00 \\
  &  & 4  & 0.14 (0.24) & 99.97 &  100.00 & 0.15 (0.24) & 99.89 \\
  &  & 9  & 0.33 (0.42) & 99.19 &  99.80 & 0.34 (0.42) & 98.14 \\
  \hline
\end{tabular} 
\end{table}

\label{lastpage}

\def\changemargin#1#2{\list{}{\rightmargin#2\leftmargin#1}\item[]}
\let\endchangemargin=\endlist 
\def\spacingset#1{\renewcommand{\baselinestretch}%
{#1}\small\normalsize} \spacingset{1}
\newcommand{\mugm}{$\mu$g/m$^3$}
\newcommand{\xiao}[1]{{\color{blue} \textbf{xiao:} #1}}

%%%%%%%%%%%%%%%%%%%%%%%%%%%%%%%%%%%%%%%%%%%%%%%%%%%%%%%%%%%%%%%%%%%%%%%%%%%%%%

\end{document}